\documentclass[aps,twocolumn,superscriptaddress,showpacs]{revtex4}
\usepackage{epsfig}
\usepackage{natbib}
\usepackage{amsmath}
\usepackage{color}
\usepackage{times}
\usepackage{psfrag}
\usepackage{epstopdf}
\newcommand{\beq}{\begin{equation}}
\newcommand{\eeq}{\end{equation}}

\begin{document}

\title{Pseudopotential  of birhythmic van der Pol type systems with correlated noise}

\author{R. Mbakob Yonkeu}
\affiliation{Laboratory of Mechanics and Materials, Department of Physics, Faculty of
Science, University of Yaound\'e I, Box 812, Yaound\'e, Cameroon.}
\author{R. Yamapi}
\email[Corresponding author:]{ryamapi@yahoo.fr
}
\affiliation{ Fundamental Physics Laboratory, Physics of Complex System group,
Department of Physics, Faculty of
 Science, University of Douala, Box 24 157 Douala, Cameroon.}
\author{G. Filatrella}
 \affiliation{Department  of Sciences and Technologies
\small and Salerno unit of CNSIM, University of Sannio, Via Port'Arsa 11,
I-82100 Benevento, Italy.}
\author{C. Tchawoua}
\affiliation{Laboratory of Mechanics and Materials, Department of Physics, Faculty of Science,
 University of Yaound\'e I, Box 812, Yaound\'e, Cameroon.}
\date{\today}


\begin{abstract}

We propose to compute the effective activation energy, usually referred to a pseudopotential or quasipotential, of a birhythmic system -- a van der Pol like oscillator -- in the presence of correlated noise.
It is demonstrated, with analytical techniques and numerical simulations, that the correlated noise can be taken into account and one can retrieve the low noise rate of the escapes. We thus conclude that a pseudopotential, or an effective activation energy, is a realistic description for the stability of  birhythmic attractors also in the presence of correlated noise.\\
 {\it To appear on Nonlinear Dynamics, 2015}

\end{abstract}

\maketitle

\newpage

\section{Introduction}

An attractor, or a stable orbit of a dynamical system, becomes metastable in the presence of noise. In fact, noise can supply enough energy to overcome a trapping potential $\Delta U,$ and therefore it is not possible to corner the system forever in a well: When fluctuations eventually occur on such a scale to overcome the barrier, the system moves outside the basin of attraction of the attractor.
The celebrated Kramers escape time
\begin{equation}
\label{Kramers}
\tau = \tau_0  e^{\frac{\Delta U}{\theta}}
\end{equation}
\noindent simply relates the energy barrier $\Delta U$ and the intensity of the uncorrelated fluctuations $\theta$. It is therefore possible to predict the  average escape time from the knowledge of the static potential (given the noise intensity).
Analogously, for oscillatory systems where two attractors coexist, noise makes the attractors metastable and the system can randomly switch from one to another.
If the two attractors are periodic orbits, characterized by different periods, the system becomes birhythmic, \emph{i.e}. it can oscillate at two different frequencies and amplitudes for the same set of parameters.
To retrieve an effective energy barrier requires in principle a search for the lowest energy trajectory among infinite escape paths \cite{Kautz94}, unless some intuition on the minimal path is available \cite{Filatrella94,Filatrella07}.
To simplify the problem, one often reverses the logic, and computes the mean escape time to retrieve the effective energy barrier \cite{Graham85,Kar04}.
The advantage of the quasipotential, apart from the elegance to enclose the stability properties in a simple energy barrier, is the possibility to extrapolate the result from relatively short escape times to rare events through Eq.(\ref{Kramers}).
Examples are the stability of the Shapiro steps employed for standard voltage \cite{Kautz88} and pinned vortices \cite{Filatrella07}.

To retrieve the activation energy from numerical simulations one can invert Eq.(\ref{Kramers}), that  amounts to estimate the slope (on a logarithmic scale) of the average escape time:

\beq
\Delta U \simeq \frac{\log\left[T(\theta_M)\right] - \log\left[ T(\theta_m)\right] }{\theta_M -\theta_m}.
\label{pseudopot}
\eeq
when $\theta_M$ and $\theta_m$ are sufficiently low.

This method has proved effective in describing the consequences of noise on a nonlinear system that exhibits two stable attractors, a modification of the van der Pol oscillator \cite{Jewett99,Yamapi10} employed to model birhythmic systems \cite{Potapov11,Abou11}.
Birhythmic systems \cite{Hounsgaard98,Geva06,Potapov11} are of interest, for example in biology, to describe the coexistence of two stable oscillatory states, a situation that can be found in some enzyme reactions \cite{li}.
Another example is the explanation of the existence of multiple frequency and intensity windows in the reaction of biological systems when they are irradiated with very weak electromagnetic fields
\cite{kaiserp,kaiser2,kaiser3,kaiser4,kaiser5,kaiser6}.
Apart the applications to Biological systems \cite{kaiserv,frohlich}, the system is particularly interesting, inasmuch it allows an analytic treatment \cite{Ghosh11}, that has also been extended to a delayed version \cite{Cheage12}.

We propose here to extend the method to the case of correlated noise.
In fact exponentially correlated noise has been extensively investigated in many systems in connection with biological oscillations  \cite{Spagnolo04,Lindner05}, we have not found references to the role of correlation in birhythmic systems. Moreover, we want to explore the possibility to extend the quasipotential approach \cite{Graham85} to the case of correlated noise.
The basic ingredient behind Eq.(\ref{pseudopot}) is an Arrhenius-type dependence of the average escape time, as in Eq.(\ref{Kramers}).
So our purpose is to show that the shape of the escape time versus the noise intensity is still Arrhenius-like in the presence of correlation.
Also by analytic treatment, we show that an effective potential naturally arises for the averaged system \cite{Anishchenko07}. 
 It has been shown that an effective trapping energy can be employed to characterize the global stability properties of the birhythmic attractors, in the presence of uncorrelated white noise \cite{Yamapi10,Cheage12}.   Our present aim is to show that the global stability properties of the attractors can be analyzed also  in the presence of exponentially correlated noise.

The paper is organized as follows. Sect. \ref{model} describes the model equation and the main properties of the birhythmic system. In Sect. \ref{effective} we discuss the results of the stochastic averaging and the properties of the pseudopotential, and that we compare with numerical simulations. Sect. \ref{conclusion} summarizes the results and the outlook.

\section{Model equations of a  self-sustained oscillator}
\label{model}

We describe in this Section the mathematical model and the main properties of a self-sustained oscillator capable to maintain, for a suitable choice of the parameters, oscillations at two different amplitudes and frequencies.
The two attractors become metastable in the presence of noise, when the system oscillates from one attractor to the other.
The random passages are in fact the main theme of this work.

\subsection{The self-sustained model}

The model considered is a van der Pol-like oscillator with a nonlinear function of higher polynomial order described by the nonlinear equation (overdots as usual stand for the derivative with respect to time)
\begin{eqnarray}
\label{eq1}
\ddot x-\mu(1-x^2+\alpha x^4-\beta x^6)\dot x+x=0
\end{eqnarray}
where $\alpha,\beta$ and $\mu$ are positive parameters that tune the nonlinearity.
Eq.~(\ref{eq1}) describes several dynamic systems, ranging from physics to engineering and biochemistry \cite{vanderpolgeneral}.
In particular Eq.~(\ref{eq1}) seems to be more appropriate for some biological processes than the classical van der Pol oscillator, as shown by Kaiser in Refs. \cite{kaiser2,kaiser3,kaiser4,kaiser5,kaiser6,kaiserp}.
When employed to model biochemical systems, namely the enzymatic-substrate reactions, $x$ in Eq.~(\ref{eq1}) is proportional to the population of enzyme molecules in the excited polar state, the quantities $\alpha$ and $\beta$ measure the degree of tendency of the system to a ferroelectric
instability, while $\mu$ is a positive parameter that tunes nonlinearity \cite{enjieu-chabi-yamapi-woafo}.
The most important feature for our purposes of the nonlinear self-sustained oscillator described by Eq.~(\ref{eq1}) is that it possesses more than one stable limit-cycle solution \cite{kaiser2}, \emph{i.e.} that it exhibits birhythmicity.

Model (\ref{eq1}) is therefore a prototype for self-sustained systems and exhibits some interesting features of nonlinear dynamical systems; for instance Refs. \cite{kaiserv,kaiserp} have analyzed the super-harmonic resonance structure and have found symmetry-breaking crisis and
intermittence.
The nonlinear dynamics and synchronization process of two such systems have been recently investigated in Refs. \cite{enjieu-chabi-yamapi-woafo,enjieu-yamapi-chabi}, while the possibility that, introducing an active control, chaos can be tamed for an appropriate choice of the coupling parameters has been considered in Ref. \cite{yamapi-nana-enjieu-2007}.

In this work we focus on model (\ref{eq1}) as a prototype for the occurrence of birhythmicity and to investigate the effects of correlation time in noise induced passages from an attractor to the other.

\subsection{Stochastic model and integral algorithm for colored noise}

In realistic environments noise affects the dynamics \cite{Tsimring14}. It is therefore conceivable that in a birhythmic systems, as an enzymatic substrate reaction with ferroelectric behavior in brain waves models (see Refs.\cite{kaiser1,frohlich,enjieu-chabi-yamapi-woafo}) such as the multi-limit-cycle van del Pol-like oscillator Eq.~(\ref{eq1}), the electrical field applied to the excited enzymes depends also on the external chemical influences  \emph{i.e.}, the flow of enzyme molecules through the transport phenomena that contain a random term.
In general, one can assume that an active oscillator of the van der Pol type (e.g., \cite{Zakharova10}) is under the environmental influence that contains a random perturbation.
Therefore, additive random excitation amounts to the Langevin version of Eq.~(\ref{eq1}), namely:
\begin{eqnarray}
\label{eq2}
\ddot x-\mu(1-x^2+\alpha x^4-\beta x^6)\dot x+x=\zeta(t).
\end{eqnarray}
We here suppose that the noise term $\zeta(t)$ is exponentially correlated \cite{middleton},
and can be generated through the following first order of stochastic differential equation:
\begin{equation}
 \label{eq3av}
   \dot \zeta=-\lambda \zeta+\lambda g_w.
 \end{equation}
The straightforward way to insert exponentially correlated colored noise of Eq.(\ref{eq3av})  is to replace the second order Eq.(\ref{eq2}) with the following system of three first order equations \cite{fox}:
\begin{subequations}
 \label{eq3}
 \begin{eqnarray}
\label{eq3a}  \dot x&=&u,  \\
\label{eq3b}  \dot u&=&\mu(1-x^2+\alpha x^4-\beta x^6)u-x+\zeta, \\
\label{eq3c}  \dot \zeta&=&-\lambda \zeta+\lambda g_w,
 \end{eqnarray}
\end{subequations}
where $g_w$ is a Gaussian white noise  whose statistical features are completely determined by the additional properties:
\begin{eqnarray}
\label{eq4} &&<g_w (t)>=0\nonumber \\
&&<g_w (t)g_w (t')>=2D \delta (t-t').
\end{eqnarray}
Thus, the priginal second order system with correlated noise is now replaced by a third order system with uncorrelated noise.
The driven noise $\zeta$ is now exponentially correlated colored noise (Ornstein-Uhlenbeck)
with the properties:
\begin{eqnarray}
\label{eq5} &&<\zeta (t)>=0\nonumber \\
&&\{<\zeta (t) \zeta (t')>\}=D \lambda \exp(-\lambda|t-t'|).
\end{eqnarray}
in which $\{...\}$ denotes averaging over the distribution of initial $\zeta_0$ values which is given by
\begin{equation}
\label{eq6}
P(\zeta_0)=\frac{1}{(2\pi D\lambda)}\exp\left[ -\frac{\zeta_0^2}{2D\lambda}\right].
\end{equation}
This secondary averaging is essential for the stationary correlation given in (\ref{eq5}).
 The parameter $D$ is the intensity of the colored noise and $\tau=\lambda^{-1}$ is the correlation time for the colored noise.
To numerically solve Eq.(\ref{eq3}), the following points are worth emphasis:
\begin{enumerate}
  \item We use the Box-Mueller algorithm \cite{knuth} to generate Gaussian white noise
  from two random numbers which are uniformly distributed on the unit interval.
  That is, to get $g_w$, we take:
\begin{subequations}
  \begin{eqnarray}
  \label{eq7}
  a,b&=&random\, numbers\, \\
    g_w&=&\left[ -4D\Delta t\ln(a) \right]^{1/2}\cos(2\pi b),
  \end{eqnarray}
\end{subequations}
  where $\Delta t$ denotes the integration step size, which was kept constant throughout the simulation. Analogously, for the initial distribution of $\zeta$, $\zeta(0)$, we use:
\begin{subequations}
  \label{eq8}
  \begin{eqnarray}
  m,n&=&random\, numbers\, \\
  \zeta(0)&=&\left[ -2D\lambda\ln(m) \right]^{1/2}\cos(2\pi n),
   \end{eqnarray}
\end{subequations}
  \item We treat Eqs. (\ref{eq3}) as three coupled equations in the variables,  $\zeta,x$ and $u$.
The integral algorithm to generate colored noise is obtained as follows: we integrate
  $\zeta$ in (\ref{eq3}) to obtain:
  \begin{equation}
  \label{eq9}
  \zeta(t)=e^{-\lambda t}\zeta(0)+\lambda \int_0^tdse^{-\lambda(t-s)}g_w(s),
  \end{equation}
  and
\begin{widetext}
  \begin{eqnarray}
  \label{eq10}
 \zeta(t+\Delta t) =
 e^{-\lambda (t+\Delta t)}\zeta(0)+\lambda \int_0^{t+\Delta t}dse^{-\lambda(t+\Delta t-s)}g_w(s).
  \end{eqnarray}

  Consequently,
   \begin{eqnarray}
  \label{eq11}
  \zeta(t+\Delta t)&&=e^{-\lambda\Delta t}\zeta(t)+
  \lambda \int_{t}^{t+\Delta t}dse^{-\lambda(t+\Delta t-s)}g_w(s) =
  e^{-\lambda\Delta t}\zeta(t)+h(t,\Delta t).
  \end{eqnarray}
\end{widetext}
Now, $h(t,\Delta t)$ is Gaussian (because $g_w$ is)
  and has zero mean (because $g_w$ does). Therefore its properties are determined by the second moment
  \begin{equation}
  \label{eq12}
  <h^2(t,\Delta t)>=D\lambda (1-e^{-2\lambda \Delta t}).
  \end{equation}
  Thus, to start the simulation, an initial value for $\zeta$ is needed and it is obtained in accord with Eq.(\ref{eq8}), and set $E=\exp\{-\lambda \Delta t\}$. The exponentially correlated noise is then obtained by the lines:
\begin{subequations}
   \begin{eqnarray}
  \label{eq13}
  a,b&=&random\, numbers\, \\
  h&=&[-2D\lambda (1-E^2)\ln(a)]^{1/2}\cos(2\pi b),\\
  \zeta_{|t+\Delta t}&=&\zeta E+h,
  \end{eqnarray}
\end{subequations}

  \item Numerical quasi-random algorithms can be used to realize both   Eq.(\ref{eq6}) and $g_w$ in Eq.(\ref{eq3}).
This yields the Euler version of the integration of Eq.(\ref{eq3}):
\begin{subequations}
   \begin{eqnarray}
  \label{eq14}
  m,n&=&random\, numbers\, \\
  \zeta&=&\left[ -2D\lambda\ln(m) \right]^{1/2}\cos(2\pi n)\\
  a,b&=&random\, numbers,  \\
  h&=&[-2D\lambda (1-E^2)\ln(a)]^{1/2}\cos(2\pi b),       \\
  x_{|t+\Delta t}&=&x+u\Delta t, \\
  u_{|t+\Delta t}&=&\nonumber  \\
   u&+&\left[\mu(1-x^2+\alpha x^4-\beta x^6)u-x+\zeta\right]\Delta t, \\
  \zeta_{|t+\Delta t}&=&\zeta E+h,
  \end{eqnarray}
\end{subequations}
\end{enumerate}
 The step size used for numerical integration is generally equal to $\Delta t=0.0001$, but in some cases we have used a smaller step. 
Such very small the time step $\Delta t$ is necessary to ensure the stability of the numerical scheme (\ref{eq14}). Moreover, the Euler integration method is only a first order method, and thus requires a very small step. We have found it convenient, for each step evaluation is very fast and the algorithm can be executed on an ordinary laptop computer.
We have also checked that averaging over as many as $200$ realizations the results converge within few percents.
A special care should be paid to estimate the escape from a basin of attraction, or in general close to an absorbing barrier, to avoid the inaccuracy due to a finite sampling of the random evolution \cite{mannella}.
We have carefully checked that the results are independent of the step size in two ways: halving the step size until stable results are reached (and with much attention to low noise intensity $D$ \cite{mannella}) and  calibrating the numerical method with a known activation barrier to retrieve the Kramer escape rate \cite{kramer}.

\begin{figure}[htb]
\centering
\begin{minipage}{9cm}
\begin{center}
\begin{picture}(140,80)
\put(-10,0.0)
{\psfig{file=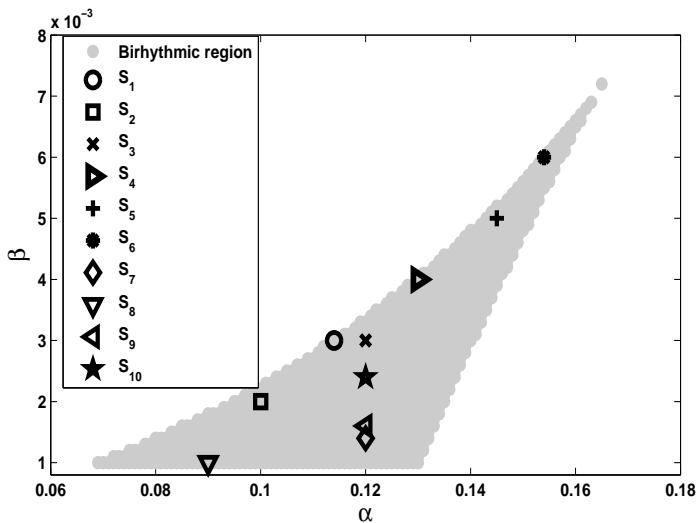,width=10.8cm,height=7.2cm,angle=0.0}}
\end{picture}
\caption {\footnotesize \it Parameter domain of the single limit cycle (white area) and three limit cycles (gray area), with $\mu=0.01$. The symbols denote the selected points of Tables \ref{identical} and \ref{different}. }
\label{fig:parameters}
\end{center}
\end{minipage}
\end{figure}

{\scriptsize
\begin{table*}
\begin{tabular}{|l|c|c|c|}
\hline
$S_i(\alpha,\beta)$& Amplitudes $A_i$ of the orbits& Frequencies $\Omega_i$ of the orbits&Periods  $P_i$ of the orbits \\
\hline
                   &$A_1$=2.3772                    & $\Omega_1$=1.0021      & $P_1=6.270$\\

$S_1(0.114;0.003)$& $A_2$=5.0264                     & $\Omega_2$=1.0011           &$P_2=6.275$ \\

                 & $A_3$=5.4667                     & $\Omega_3$=1.0231      &$P_3=6.141$ \\
\hline
                        & $A_1$=2.3069           & $\Omega_1$=0.9870      &$P_1=6.366$ \\

$S_2(0.1;0.002)$     & $A_2$=4.8472                    & $\Omega_2$=1.0001             &$P_2=6.275$ \\

                        & $A_3$=7.1541                 & $\Omega_3$=0.97123     &$P_3=6.468$ \\
\hline
                        & $A_1$=2.4269                   & $\Omega_1$=0.9850       &$P_1=6.379$ \\

$S_3(0.12;0.003)$     & $A_2$=4.2556                  & $\Omega_2$=0.9990            & $P_2=6.289$ \\

                        & $A_3$=6.3245             & $\Omega_3$=0.9865       &$P_3=6.369$ \\
\hline
                        & $A_1$=2.4903               & $\Omega_1$=1.0002   &$P_1=6.282$ \\

$S_4(0.13;0.004)$     & $A_2$=4.4721                    & $\Omega_2$=1.0001        & $P_2=6.282$\\

                        & $A_3$=5.0791               & $\Omega_3$=0.9991       & $P_3=6.289$\\
\hline
                         & $A_1$=2.6605              & $\Omega_1$=1.0002      & $P_1=6.282$\\

$S_5(0.145;0.005)$    & $A_2$=3.8305                    & $\Omega_2$=1.0001          & $P_2=6.282$\\

                        & $A_3$=4.9643                 & $\Omega_3$=1.0005    &$P_3=6.280$ \\
\hline
                        & $A_1$=2.7864                & $\Omega_1$=0.9992     &$P_1=6.288$ \\

$S_6(0.154;0.006)$     & $A_2$=3.8821                   & $\Omega_2$=1.0001          & $P_2=6.282$\\
                        & $A_3$=4.2698                & $\Omega_3$=1.0002      & $P_3=6.282$\\
\hline
\end{tabular}
{\it \caption{ \small  Characteristics of the limit cycles when the two stable frequencies are about equal (\emph{i.e.} $\Omega_1 \simeq \Omega_3$). All data refer to the case $\mu=0.01$.
\label{identical} }  }
\end{table*}
}
\subsection{Birhythmic properties}
\label{birhythmic}

In this Section we summarize the dynamical attractors of the
modified  van der Pol model (\ref{eq1}) without random excitation.
The periodic solutions of Eq.~(\ref{eq1}) can be approximated by
\begin{eqnarray}
\label{eq15} x(t)=A\cos \Omega t.
 \end{eqnarray}
We recall that approximated analytic estimates of the amplitude $A$ and the frequency  $\Omega $ have been derived in Ref. \cite{enjieu-chabi-yamapi-woafo}, and it has been found that the amplitude $A$ is independent of the coefficient $\mu$, that only enters in the frequency $\Omega_i$ of the orbits.
The parameters $\alpha$ and $\beta$ determine if the modified van der Pol system Eq. (\ref{eq1}) posses one or three limit cycles
(When three limit cycles are obtained, two of them are stable and one is unstable).
The unstable limit cycle represents the separatrix between the basins of attraction of the two stable limit cycles at different frequencies (and hence the system is birhythmic).
We show in Fig.\ref{fig:parameters} the  bifurcation lines that contour the
region of existence of birhythmicity in the two parameters in the  phase
space ($\beta$-$\alpha$)
\cite{enjieu-chabi-yamapi-woafo,enjieu-yamapi-chabi}.
 The bifurcation border of the shadowed area on the left denotes the passage from a single limit cycle to three limit cycles, while the right border denotes the reverse passage from three limit cycles to a single solution. At the conjunction, a codimension-two bifurcation, or cusp \cite{Anishchenko07}, appears. The first bifurcation encountered increasing $\alpha$ corresponds to the saddle-node bifurcation of the outer, or larger amplitude cycle, while the second bifurcation occurs in correspondence of a saddle-node bifurcation of the inner, or smaller amplitude, cycle. The two frequencies associated to the limit cycles are very similar close to the lowest $\alpha$ bifurcation (Table \ref{identical}) and clearly distinct at the highest $\alpha$ bifurcation line (Table \ref{different}).

\begin{table*}
\begin{tabular}{|l|c|c|c|}
\hline
$S_i(\alpha,\beta)$& Amplitudes $A_i$ of the orbits& Frequencies $\Omega_i$ of the orbits&Periods $P_i$ of the orbits\\
\hline
                   &$A_1$= 2.491378                 & $\Omega_1$=1.00210& $P_1=6.2698$\\

$S_7(0.12;0.0014)$& $A_2$=3.52558                    & $\Omega_2$=0.99994     &$P_2=6.2834$ \\

                 & $A_3$=10.88605                     & $\Omega_3$=0.57300 &$P_3=10.9651$ \\
\hline
                        & $A_1$= 2.26969              & $\Omega_1$=0.99986 &$P_1=6.2839$ \\

$S_8(0.09;0.001)$     & $A_2$= 4.59373                & $\Omega_2$=0.99974         &$P_2=6.2847$ \\

                        & $A_3$= 10.85109            & $\Omega_3$=0.68107&$P_3=9.2252$ \\
\hline
                        & $A_1$=  2.48185           & $\Omega_1$=1.00034 &$P_1=6.2809$ \\

$S_9(0.12;0.0016)$     & $A_2$=  3.58637             & $\Omega_2$=0.99993        & $P_2=6.2845$ \\

                        & $A_3$=   10.04878           & $\Omega_3$= 0.72935&$P_3=9.2352$ \\
\hline
                        & $A_1$=  2.44836             & $\Omega_1$=0.98930 &$P_1=6.3510$ \\

$S_{10}(0.12;0.0024)$     & $A_2$=  3.88657         & $\Omega_2$=0.99989       & $P_2=6.2837$\\

                        & $A_3$=  7.67464         & $\Omega_3$=0.86703 & $P_3=7.2466$\\
\hline
\end{tabular}

{\it \caption{ \small Characteristics of the limit cycles when the two stable frequencies are about equal (\emph{i.e.} $\Omega_1 \neq \Omega_3$). All data refer to the case $\mu=0.01$.
\label{different}}  }
\end{table*}

\begin{figure*}
\begin{center}
\includegraphics[height=5cm,width=7cm]{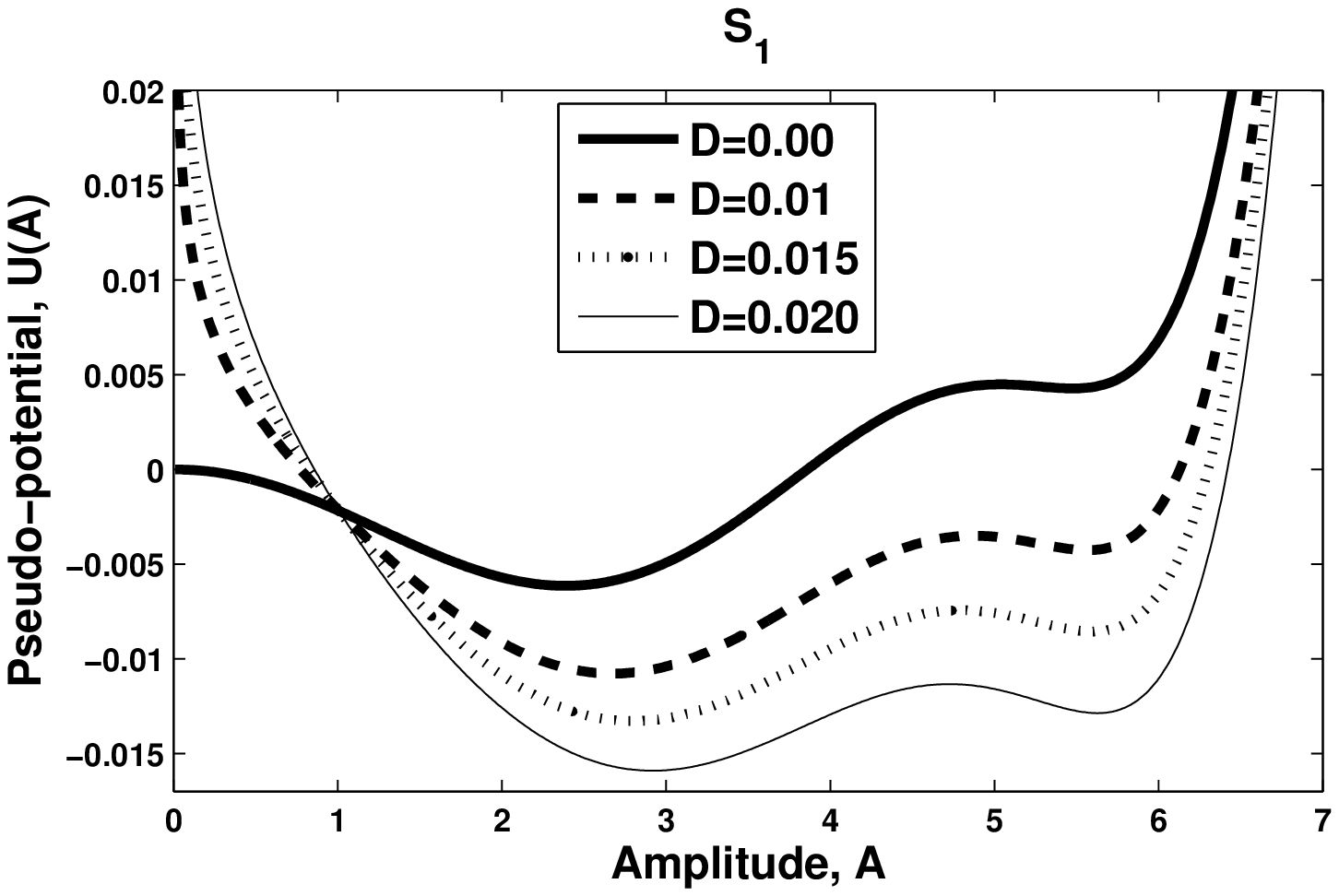} \hspace{1cm}
\includegraphics[height=5cm,width=7cm]{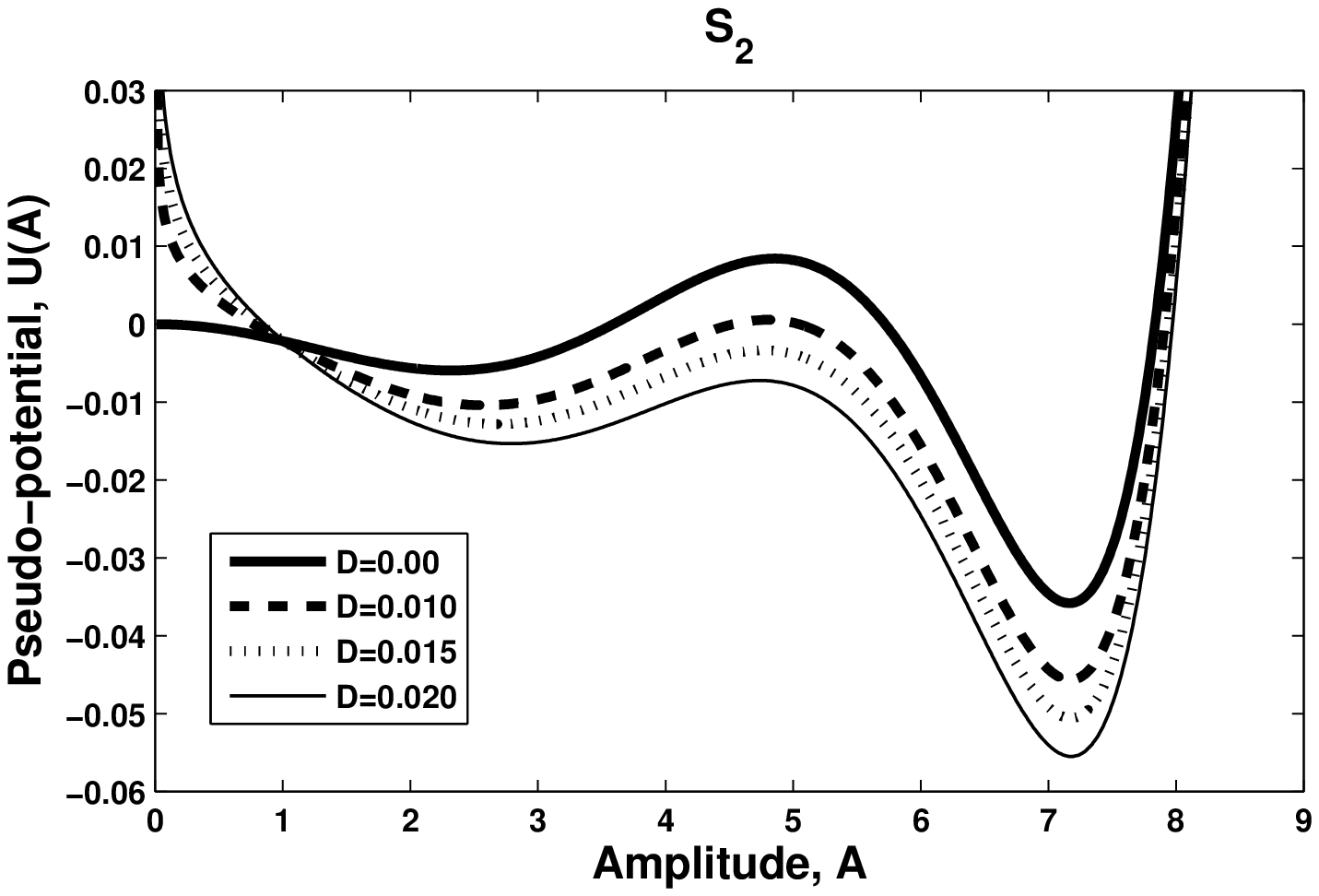}\\
\includegraphics[height=5cm,width=7cm]{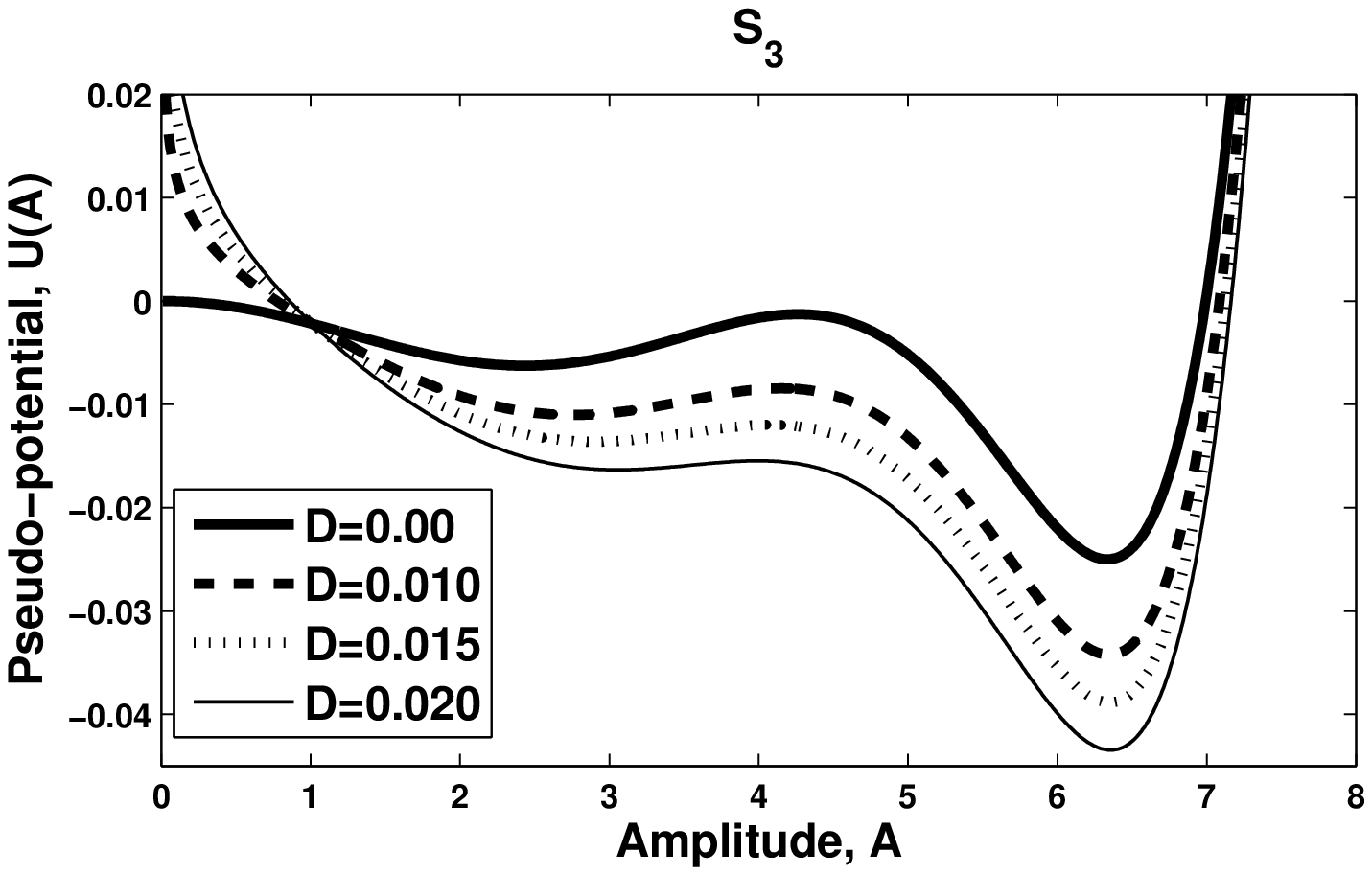} \hspace{1cm}
\includegraphics[height=5cm,width=7cm]{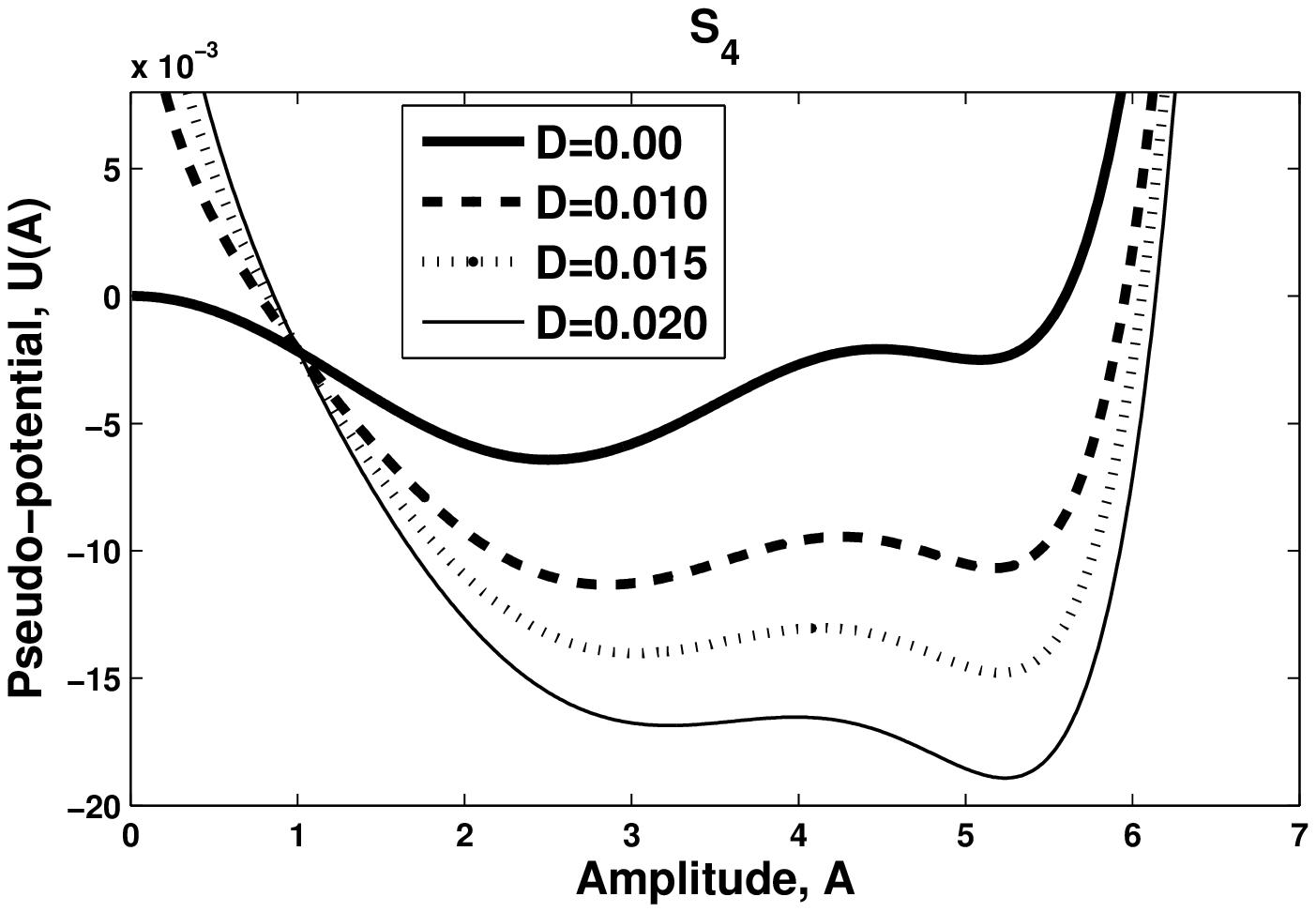}\\
\includegraphics[height=5cm,width=7cm]{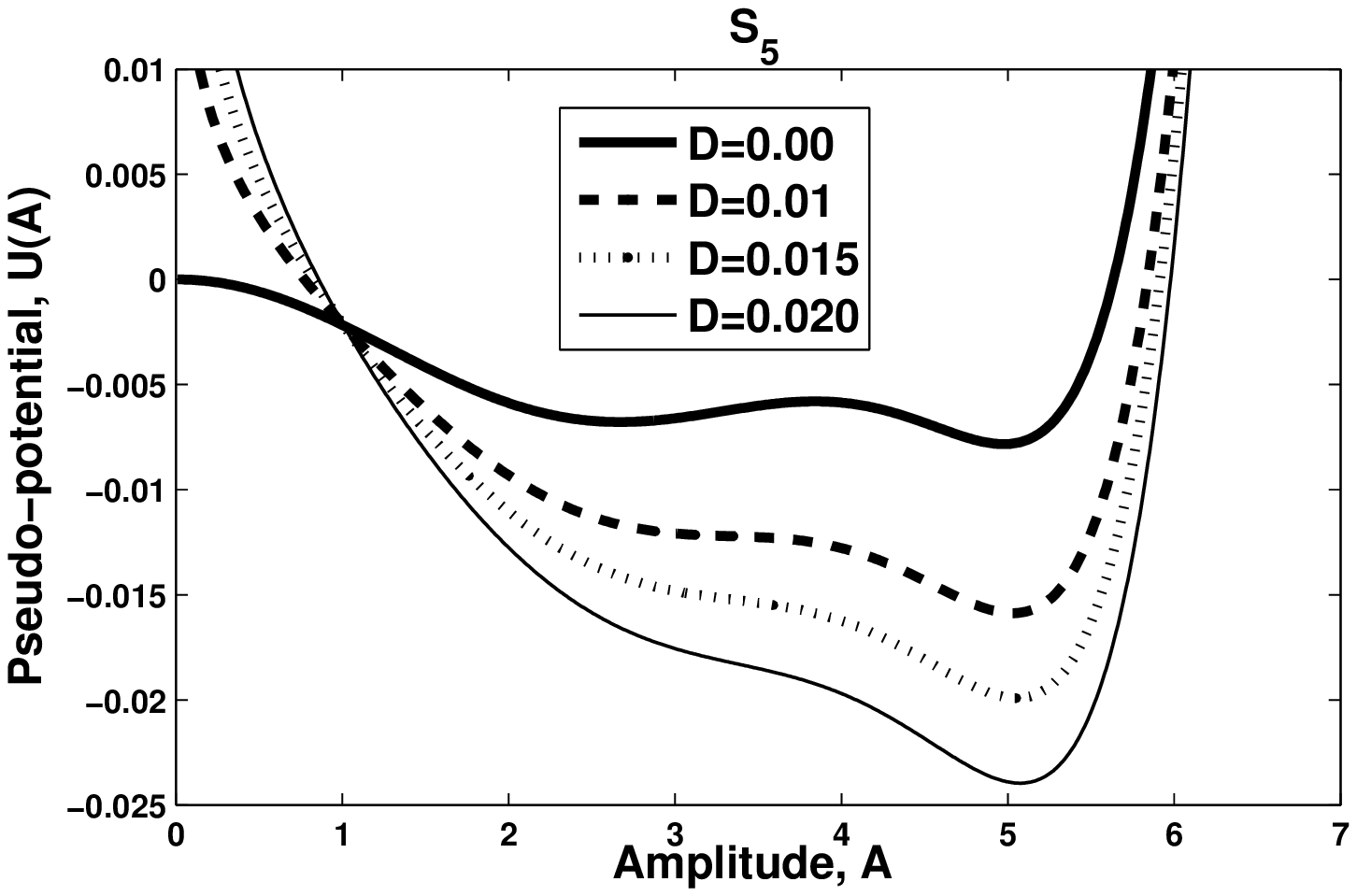} \hspace{1cm}
\includegraphics[height=5cm,width=7cm]{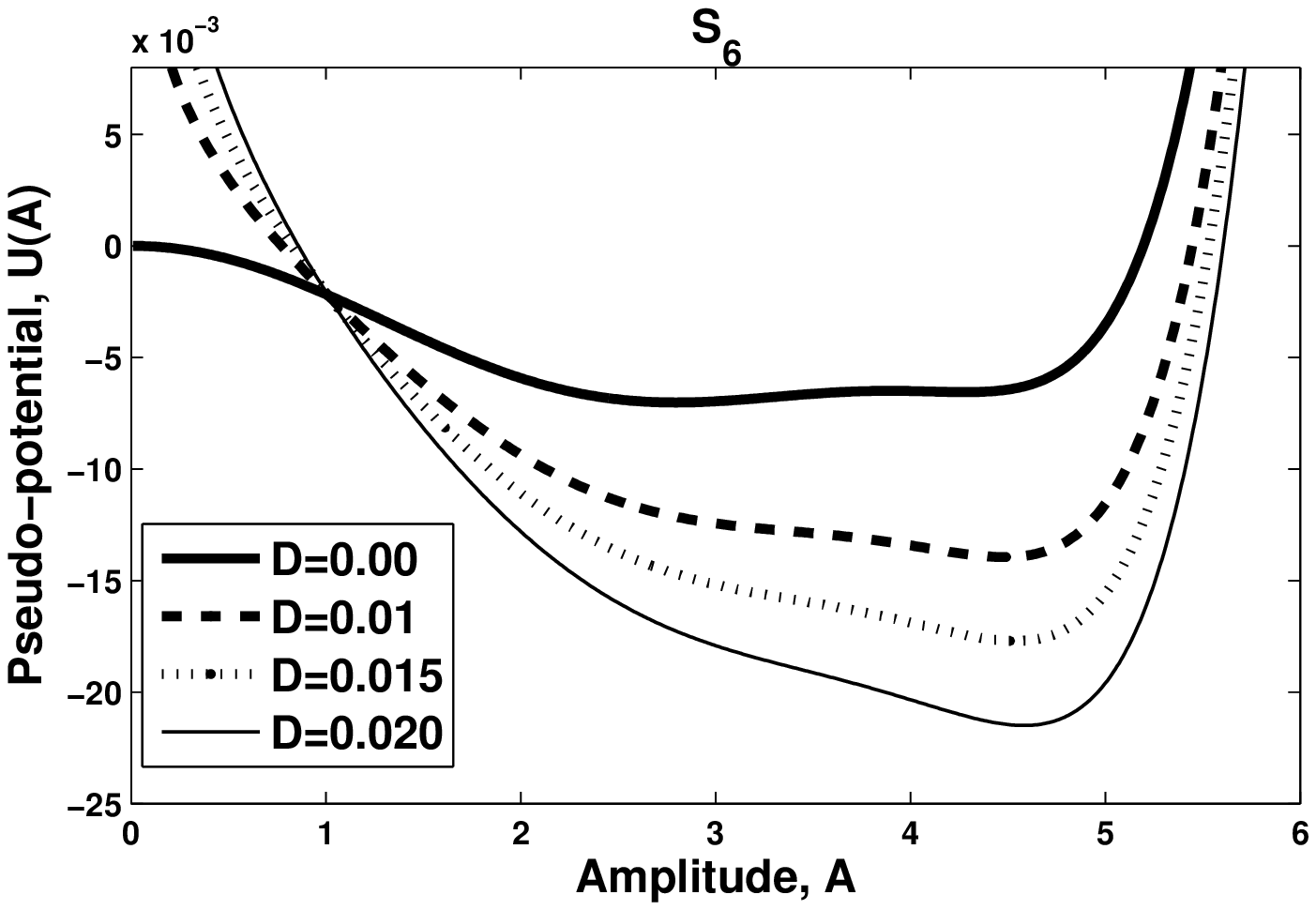}\\
\caption{\it Effects of the amplitude of the noise $D$ on the pseudopotential $U(A)$, Eq.(\ref{eq20}), for the set of parameters $\alpha$ and $\beta$ of $S_i(i=1,2,3,4,5,6)$ in Table \ref{identical}, where the frequencies of the two attractors are almost identical. The correlation reads $\tau=0.01$.
\label{fig:effects1}
}
\end{center}
\end{figure*}

\section{Effective pseudo-potential in presence of correlated noise}
\label{effective}

In this Section we propose an analytical treatment of the Langevin Eq.(\ref{eq1}).
In Subsect. \ref{analytical} we sketch the derivation of an effective Langevin equation for the averaged system \cite{Anishchenko07}.
At variance with previous derivations we include a finite correlation time, to prove that (in the limit of the stochastic averaged system) it amounts to a nonlinear scaling of the noise intensity. In Subsect. \ref{influence} we show how this rescaling influences the potential.

\subsection{Analytical estimation of the pseudo-potential}
\label{analytical}

This Subsection deals with an analytical investigation of the pseudo-potential associated to the equation of the self-sustained system (\ref{eq1}).
First, we treat Eq.(\ref{eq2}) with the average method \cite{b1,b2} to find an
equation for the amplitude that amounts  to an effective Langevin equation with
an effective associated pseudo-potential. Let us rewrite Eq.(\ref{eq2}) in the following form:
\begin{equation}
\label{eq1a}
    \ddot x+x=\mu f(x,u)+\varsigma(t)
\end{equation}
where
$$
    f(x,u)=(1-x^2+\alpha x^4-\beta x^6)u
$$
In the quasiharmonic regime, assuming that
 the noise intensity is small, one adopts the
  change of variables:
\begin{subequations}
\label{eq16}
\begin{eqnarray}
 x(t)&=&A(t)\cos(t+\varphi(t)),\\
 u(t)&=&\dot x(t)=- A(t)\sin(t+\varphi(t)).
 \end{eqnarray}
 \end{subequations}
Substituting Eq.(\ref{eq16}) into Eq.(\ref{eq1a}),
and setting $\theta \equiv t+\varphi$  we obtain
\begin{subequations}
 \begin{eqnarray}
\label{eq17}
 \dot{A}&=&-(\mu f(x,u)+\varsigma(t))sin\theta\\
 \dot{\varphi}&=&-\frac{1}{A}(\mu f(x,u)+\varsigma(t))cos\theta.
 \end{eqnarray}
 \end{subequations}
\begin{widetext}
Averaging over the period of oscillations, we obtain the following stochastic  equations for the slow (on a scale of $T=2\pi$)  varying amplitude $A(t)$ and phase $\varphi(t)$  \cite{b3}:
 \begin{subequations}
\label{5}
\begin{eqnarray}
    \dot{A}&=&-\frac{1}{2 \pi }\int_{0}^{2\pi}(\mu f(x,u)+\varsigma(t))sin\theta )d\theta=
    \frac{\mu A }{128}[64-16A^{2}+8\alpha A^{4}-5\beta A^{6}]-\frac{1}{2 \pi}\int_{0}^{2\pi}\varsigma(t)sin\theta d\theta\quad \quad \,\\
\dot{\varphi}&=&-\frac{1}{2 \pi A} \int_{0}^{2\pi}(\mu f(x,u)+\varsigma(t))\cos\theta)d\theta =
 -\frac{1}{2 \pi A} \int_{0}^{2\pi}\varsigma(t)cos\theta d\theta
\label{phidot}
\end{eqnarray}
\end{subequations}
\end{widetext}
the archetypal source for colored noise consists of an exponentially correlated process
 given by a Gauss Markov process $ \varsigma(t)$:
\begin{equation}\label{7}
    \dot{\varsigma}=-\frac{1}{\tau} \varsigma + \frac{\sqrt{D}}{\tau}g_{w}
\end{equation}
\begin{equation}\label{8}
    <\varsigma(t) \varsigma(s)> \equiv S(t-s).
\end{equation}
 Its spectrum thus assumes a Lorentzian form, i.e
\begin{equation}\label{9}
    S(a)=\int_{-\infty}^{+\infty}S(t)exp(iat)dt=S(-a)=\frac{2D}{1+\tau^{2}a^{2}}
\end{equation}
In the following discussion, we restrict the discussion to Gaussian processes with exponential correlation functions unless stated otherwise.
Let us underline the following properties of integration of the trigonometric function in the stochastic processes \cite{b3}.
If $\cos (k(t))$ is the periodic function with the period $T$, $l$ a constant and
 $W$  a Wiener processe, then
\begin{subequations}
\begin{eqnarray}
  \int_l^{l+T}\varsigma(t)\cos (k(t))dt &=& T \sqrt{\frac{D}{1+\tau^2}}\dot W,\\
  \int_l^{l+T}\dot \varsigma (t)\cos (k(t))dt&=& T \sqrt{\frac{D}{1+\tau^2}}\dot W.
\end{eqnarray}
\end{subequations}
The integral terms of Eqs.(\ref{5}) become
\begin{eqnarray}
\label{2}
& &   -\frac{1}{2\pi}\int_{0}^{2\pi}\varsigma(t)sin\theta d\theta =\nonumber \\
& &=    \frac{1}{2\pi}\left[ \varsigma(t)\cos \theta\right]_0^{2\pi}+
   \frac{1}{2\pi}\int_0^{2\pi}\dot \varsigma (t)\cos\theta d\theta = \nonumber\\
& & =\frac{\tilde{D}}{2(1+\tau^{2})}+
   \sqrt{\frac{D}{1+\tau^2}}\dot{W}_{1}
   \end{eqnarray}
 where $\tilde{D}\equiv D/A$ is a dimensionless noise intensity and
\begin{equation}\label{3}
   -\frac{1}{2\pi}\int_{0}^{2\pi}\varsigma(t)cos\theta d\theta=
   \sqrt{\frac{D}{1+\tau^{2}}}\dot {W}_{2}.
\end{equation}
Here $W_1(t )$ and $W_2(t )$ represent independent normalized Wiener processes.
The effects of colored noise can be dealt with by an effective Langevin equation. Using Eqs.(\ref{5}) and after integration and normalization, we obtain the following equations:
\begin{figure}
\begin{center}
\includegraphics[height=5cm,width=7cm]{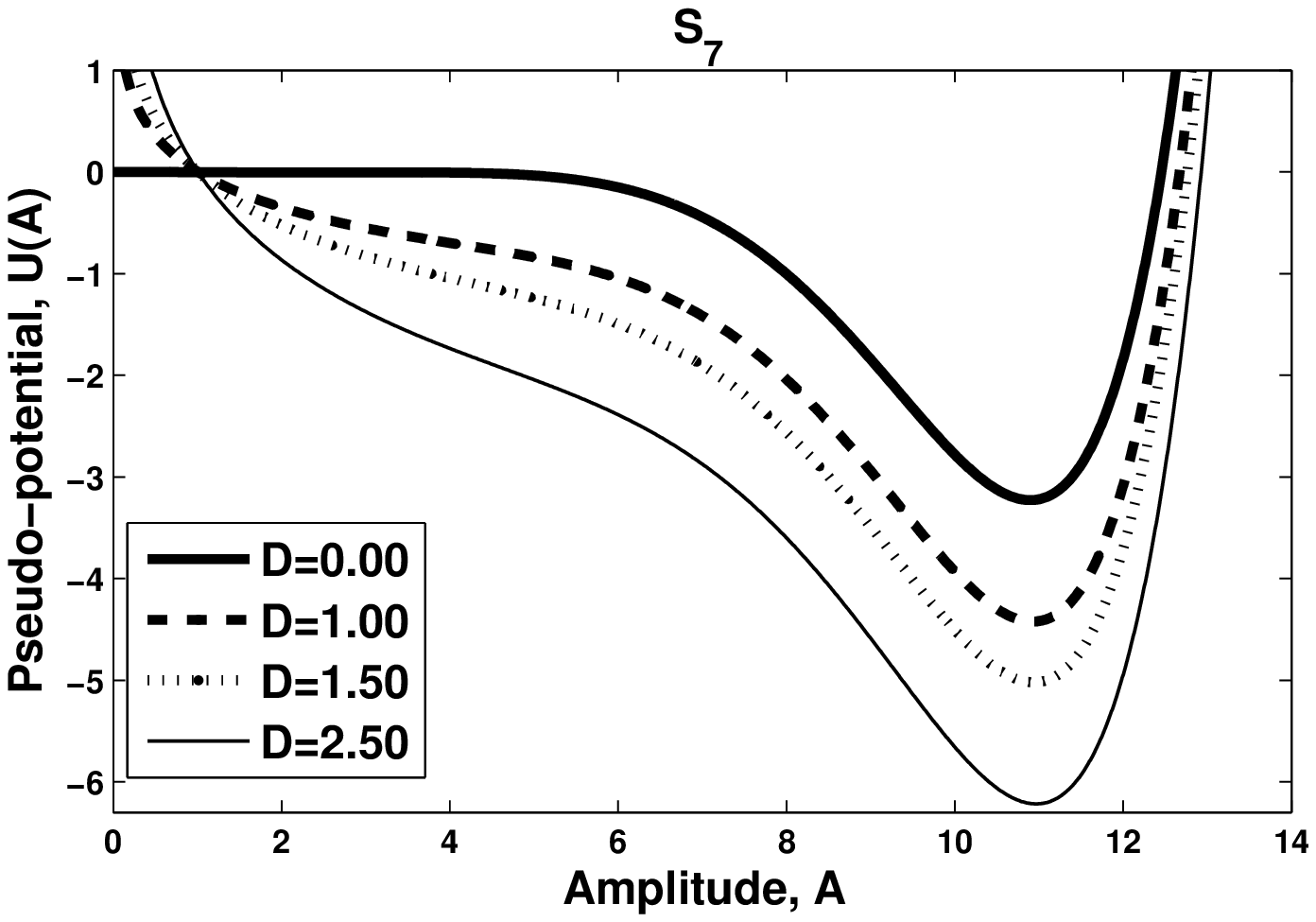}
\includegraphics[height=5cm,width=7cm]{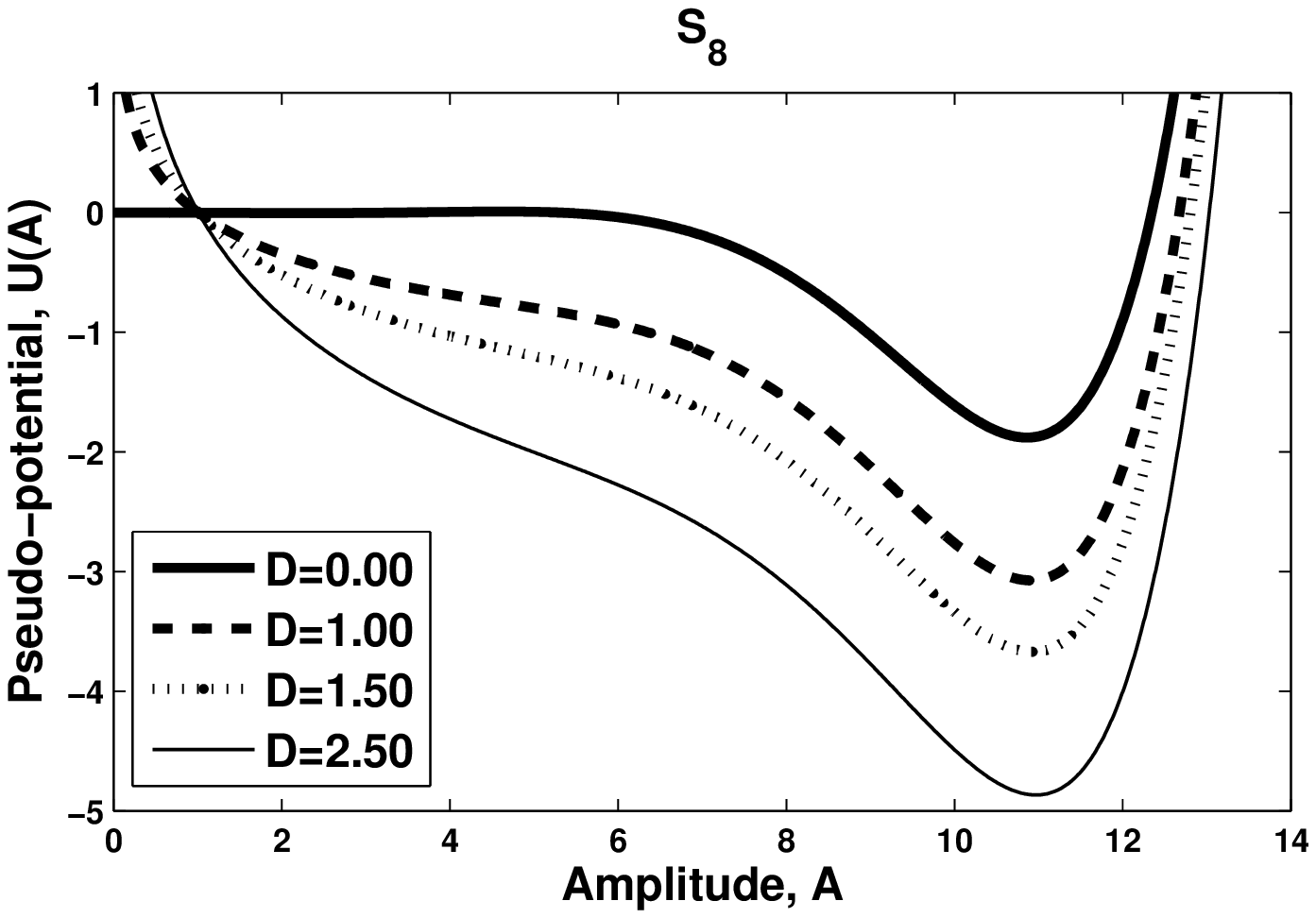}\\
\includegraphics[height=5cm,width=7cm]{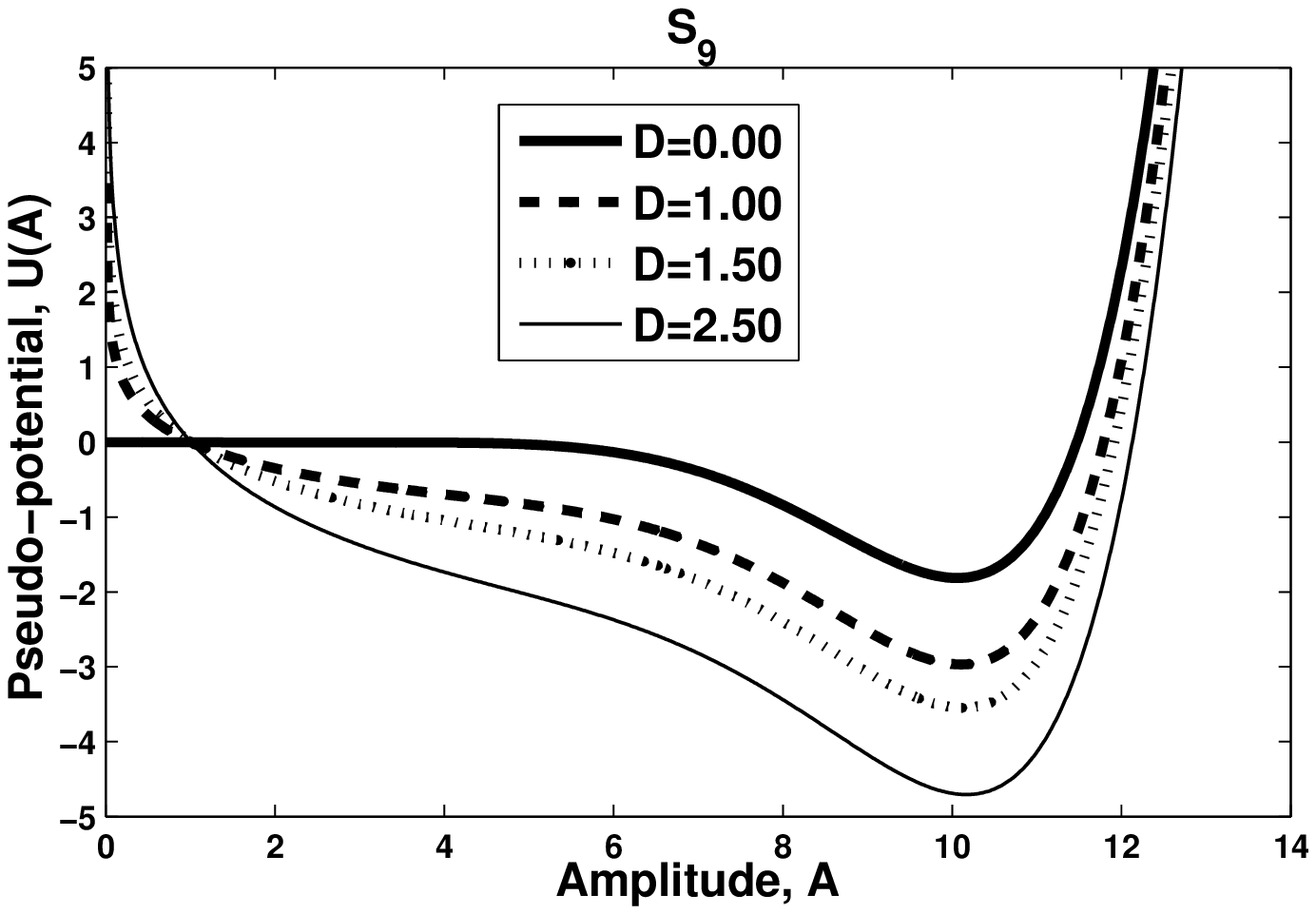}
\includegraphics[height=5cm,width=7cm]{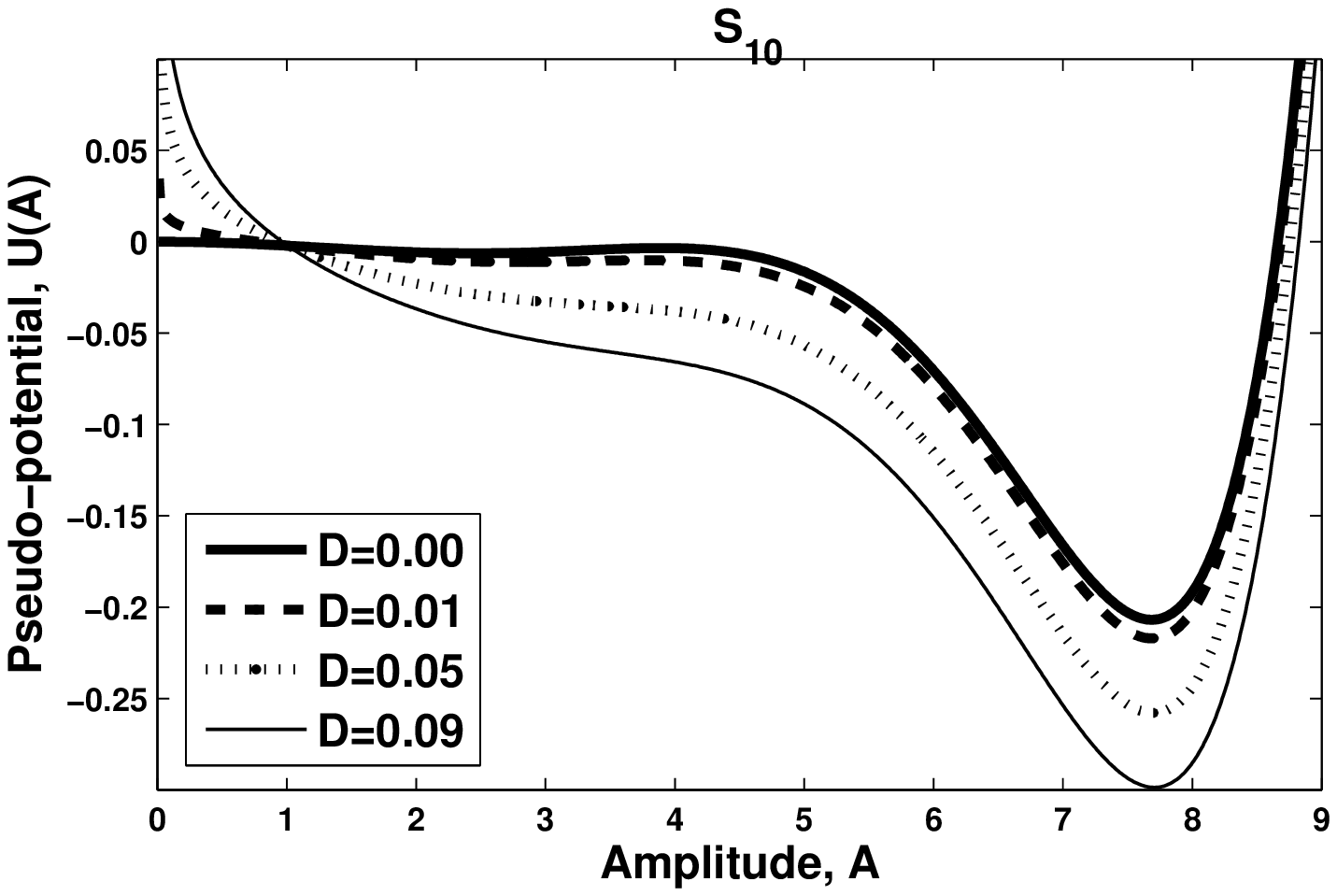}\\
\caption{\it
 Effects of the amplitude of the noise $D$ on the pseudopotential $U(A)$, Eq.(\ref{eq20}), for the set of parameters $\alpha$ and $\beta$ of $S_i(i=7,8,9,10)$ in Table \ref{different}, where the frequencies of the two attractors are different. The correlation reads $\tau=0.01$.
}
\label{fig:effects2}
\end{center}
\end{figure}
   \begin{eqnarray}
   \label{eq18}
    \dot{A}&=&\frac{\mu A }{128}[64-16A^{2}+8\alpha A^{4}-5\beta A^{6}]+\frac{D}{2A(1+\tau^{2})} \nonumber \\
 & &    +\sqrt{\frac{D}{1+\tau^{2}}}\,\dot{W}_1(t)\nonumber\\
    \dot{\varphi}&=&  \frac{1}{A}\sqrt{\frac{D}{1+\tau^{2}}}\dot{W}_{2}(t).
   \end{eqnarray}

The derivation is based on the assumption that the amplitude $A(t)$ does not change much on a cycle, i.e. over a time scale $2\pi$.
This poses a serious difficulty for the cases (Table \ref{different}) where the two frequencies in the system are different, while it is satisfied to a good approximation when the two frequencies are similar (Table \ref{identical}).
However, we apply the method assuming that the change of variable (\ref{eq16}) is valid.
From Eq. (\ref{eq17}), for $D= 0$, the nonzero amplitude of limit cycle satisfies the following equation
\begin{equation}
\label{eq18}
\frac{5\beta}{64}A^6-\frac{\alpha}{8}A^4+\frac{1}{4}A^2-1=0
\end{equation}
which coincides with the deterministic amplitude equation derived in \cite{Cheage12}.
The amplitude equation gives the range of existence of birhythmicity, that strongly depends on nonlinear parameters $\alpha$ and $\beta$, as it appears in Fig. \ref{fig:parameters}.
Analogously to Refs. \cite{Yamapi10,Cheage12}, the effective Langevin equation (\ref{eq17}) amounts to the Brownian motion of particle in a double well potential, that can be approximated in the potential form:
\begin{equation}
\label{eq19}
    \dot{A}=-\frac{dU(A)}{dA}+\sqrt{\frac{D}{1+\tau^{2}}}\,\dot{W}_1(t)
\end{equation}
where the effective pseudo-potential $U(A)$ is given by
\begin{equation}
\label{eq20}
    -U(A)=\frac{\mu }{128}[32 A^{2}-4A^{4}+\frac{4}{3}\alpha A^{6}-\frac{5}{8}\beta A^{8}]+\frac{D \ln(A)}{2(1+\tau^{2})}
\end{equation}

\begin{figure}
\begin{center}
\includegraphics[height=5cm,width=7cm]{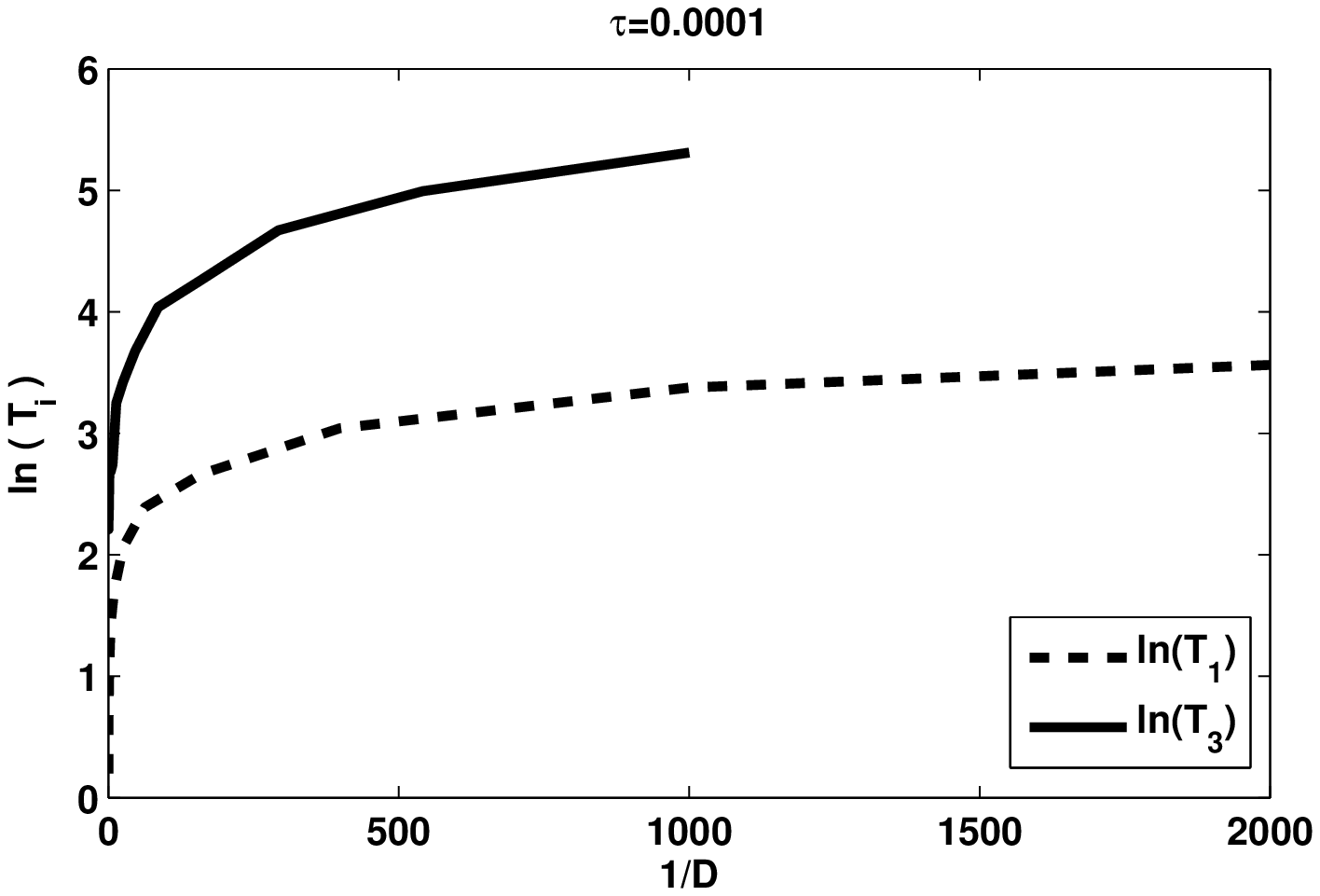}
\includegraphics[height=5cm,width=7cm]{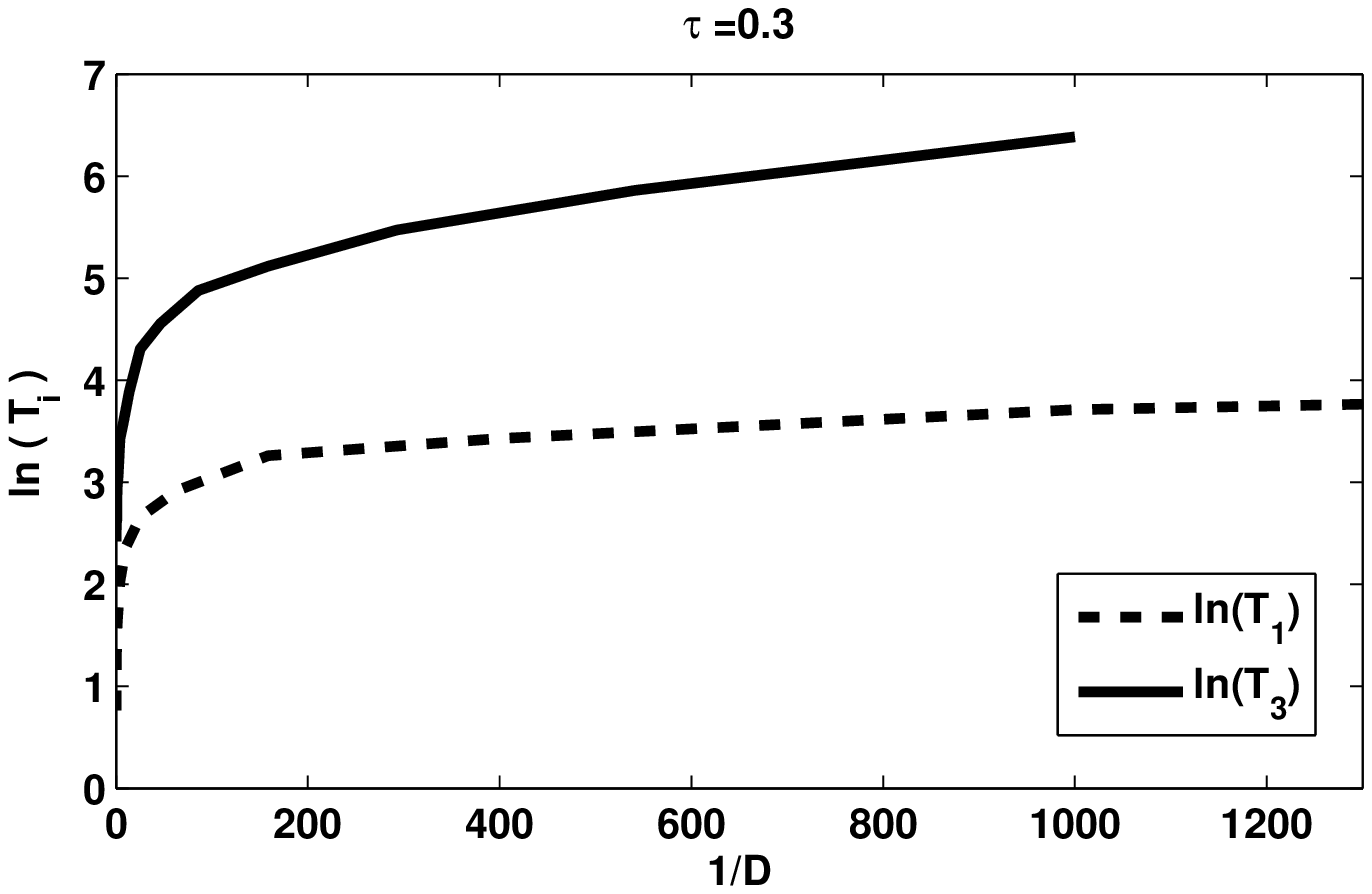}\\
\includegraphics[height=5cm,width=7cm]{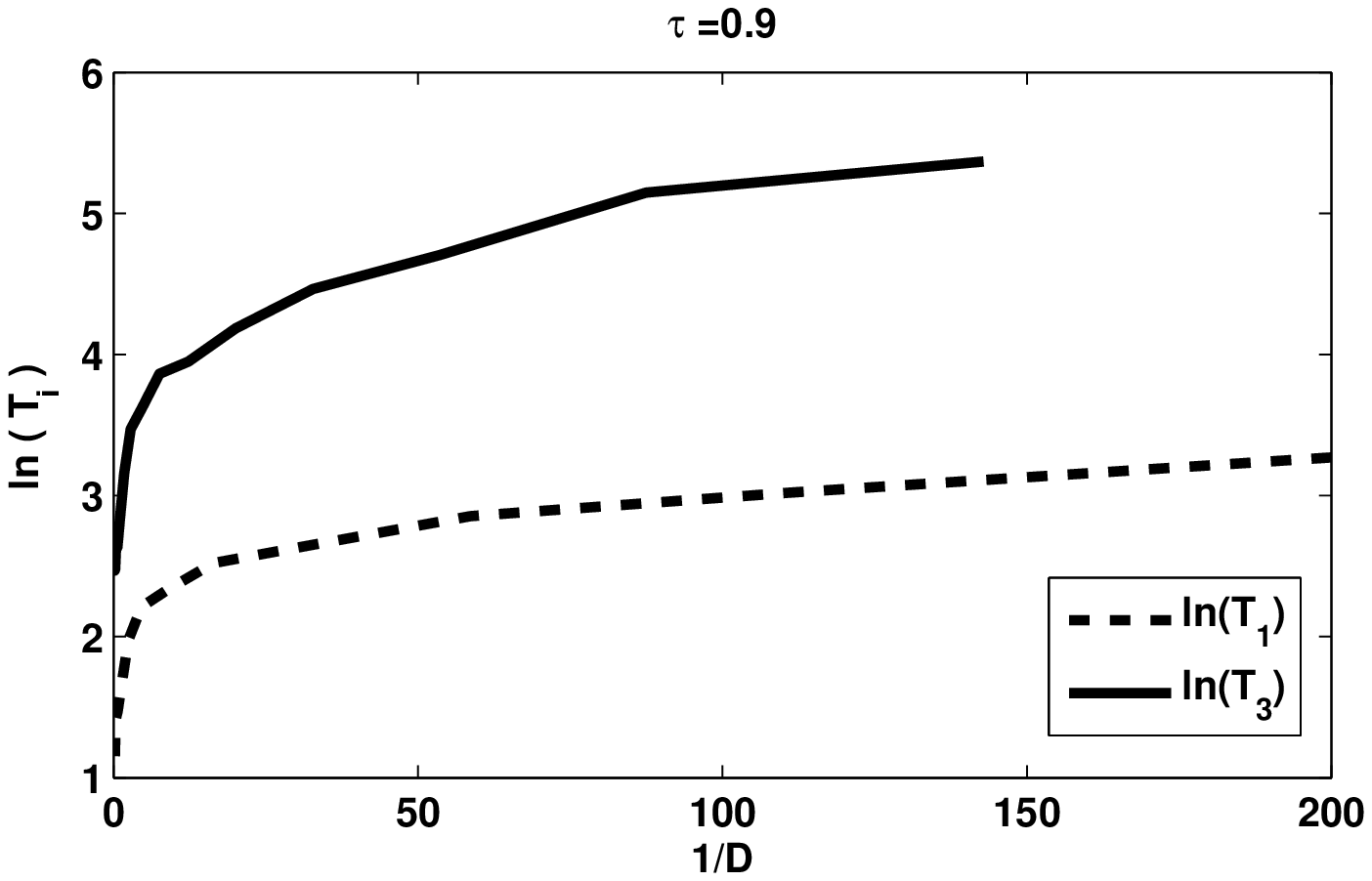}
\includegraphics[height=5cm,width=7cm]{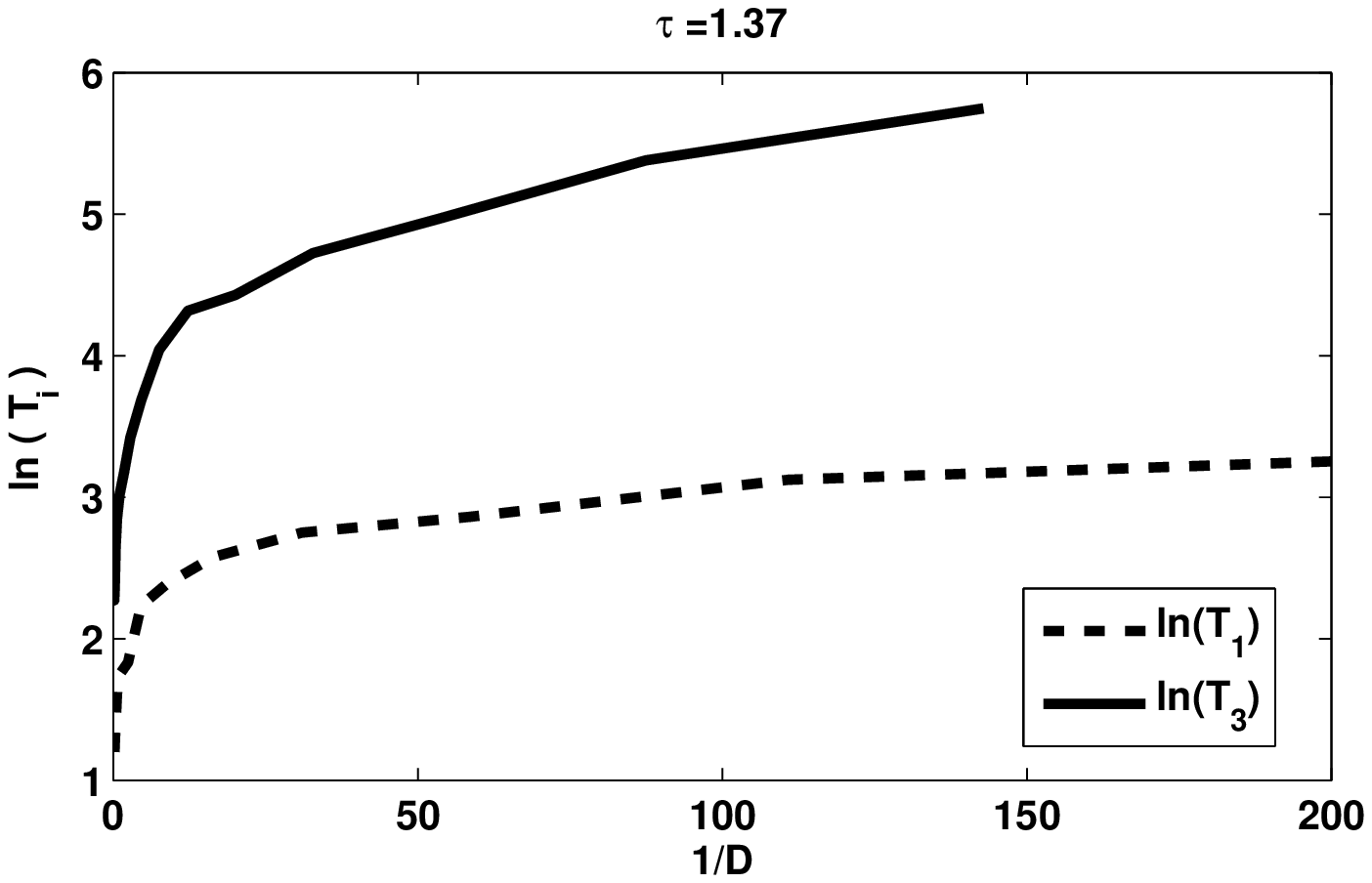}
\caption{\it
 Effect of the correlation time $\tau$ on the variation of the logarithmic average of the escape time from an attractor $A_1\Longleftrightarrow A_3$ versus 1/D, for the sets of parameters $S_1,$ shown in Table \ref{identical} (where the frequencies of the two attractors are nearly identical). The solid line ($T_3$) refers to escape from the outer circle, while the dashed line ($T_1$) refers to escapes from the inner circle. }
\label{fig:correlation1}
\end{center}
\end{figure}

\begin{figure}
\begin{center}
\includegraphics[height=5cm,width=7cm]{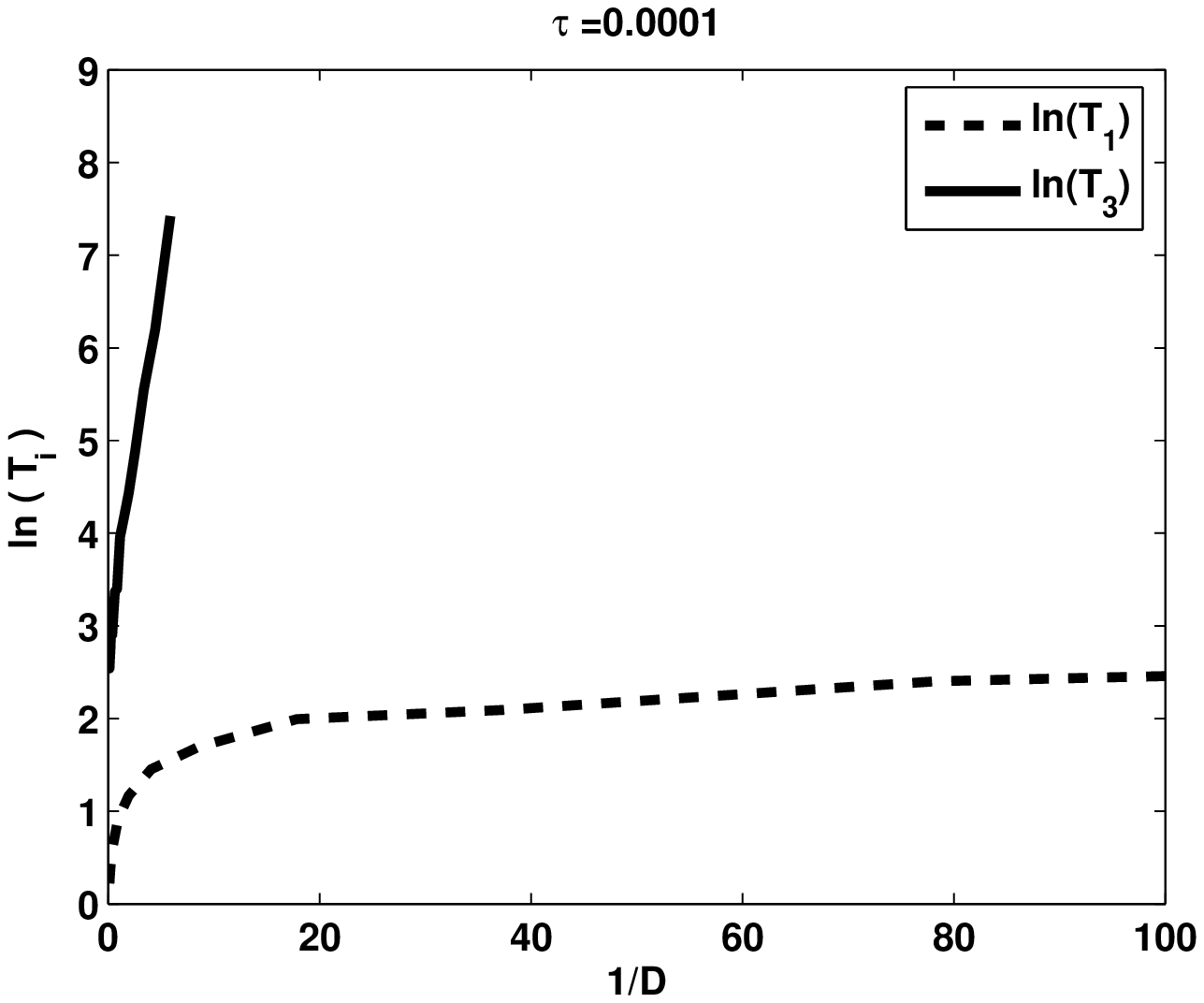}
\includegraphics[height=5cm,width=7cm]{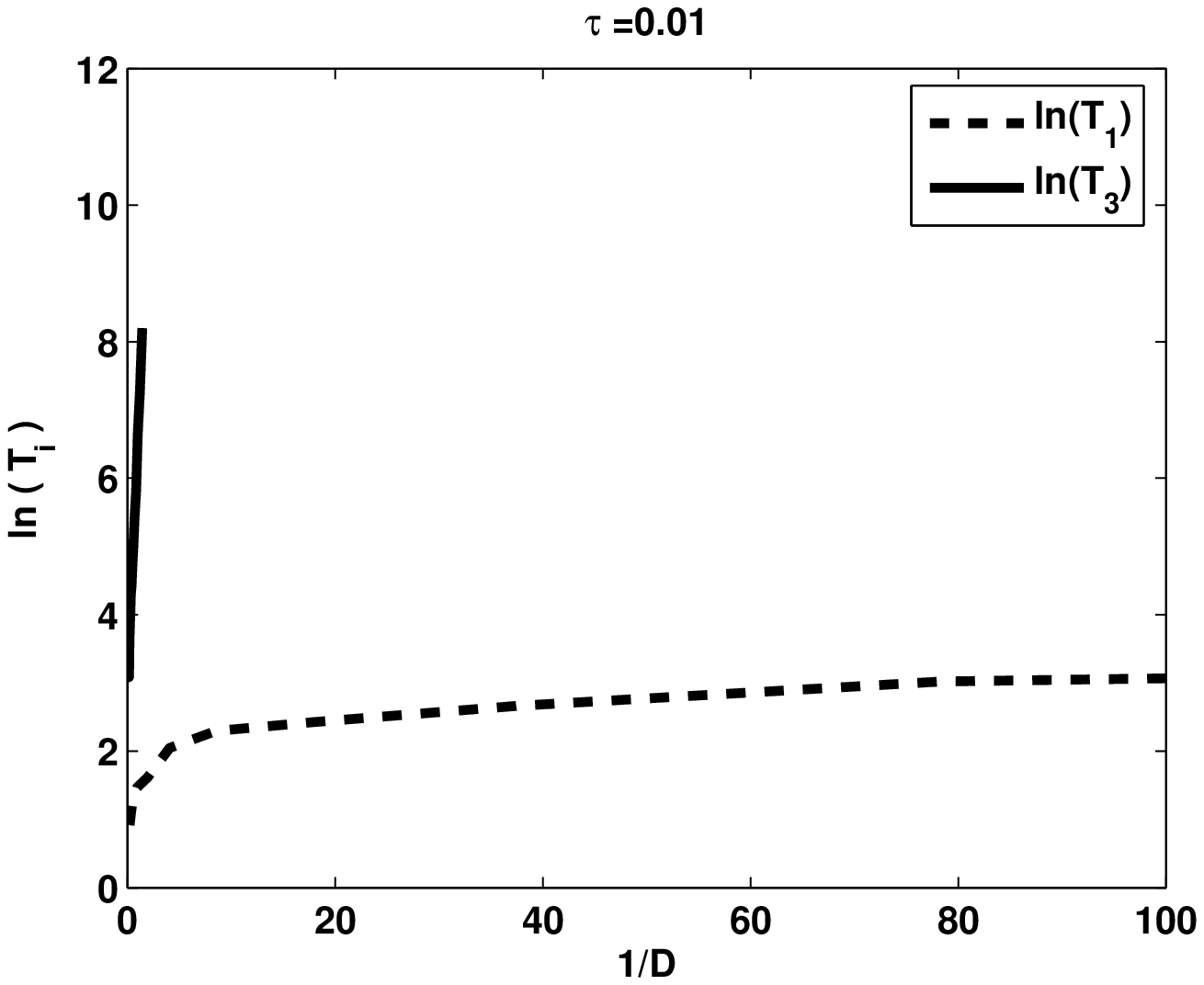}\\
\includegraphics[height=5cm,width=7cm]{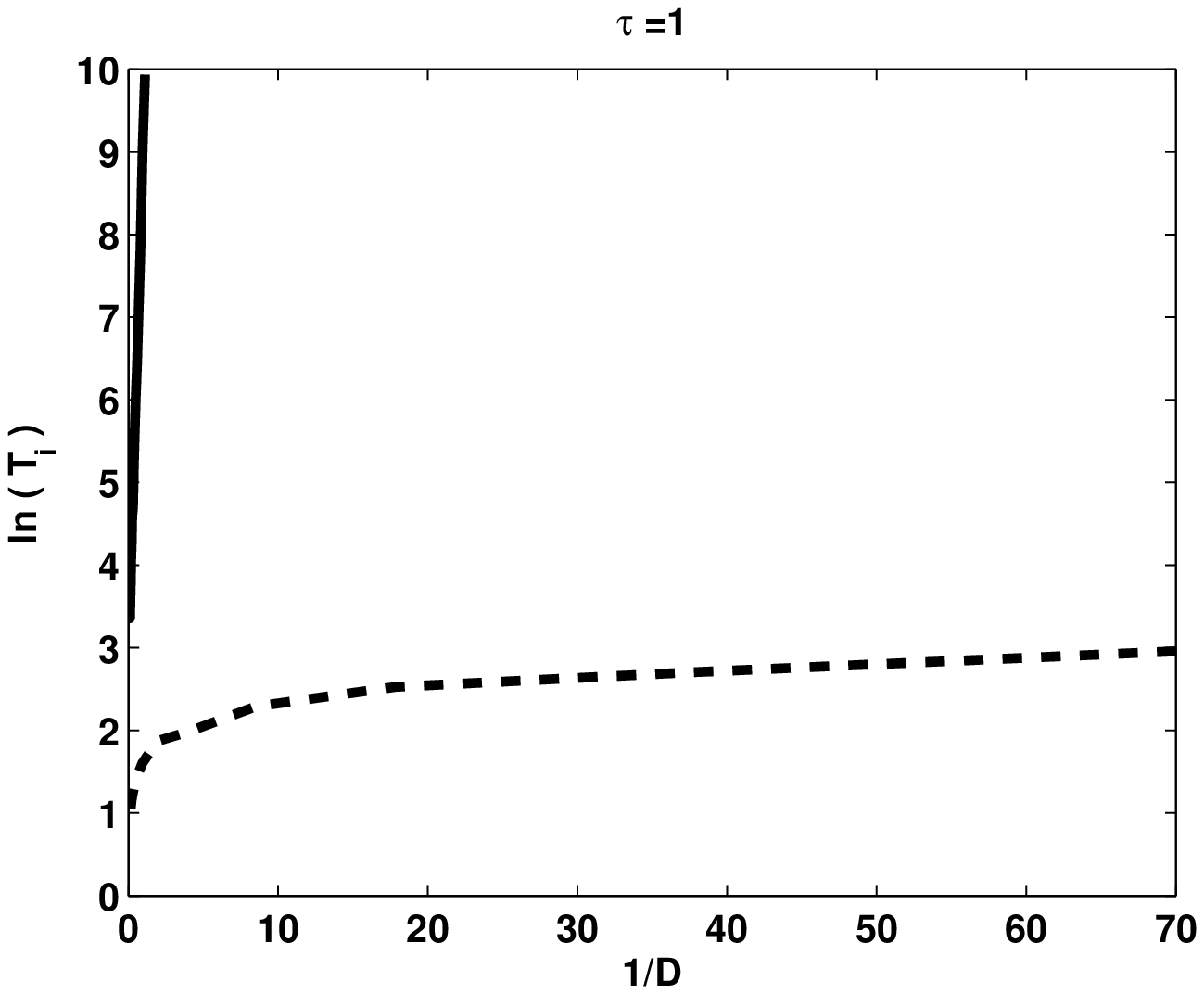}
\includegraphics[height=5cm,width=7cm]{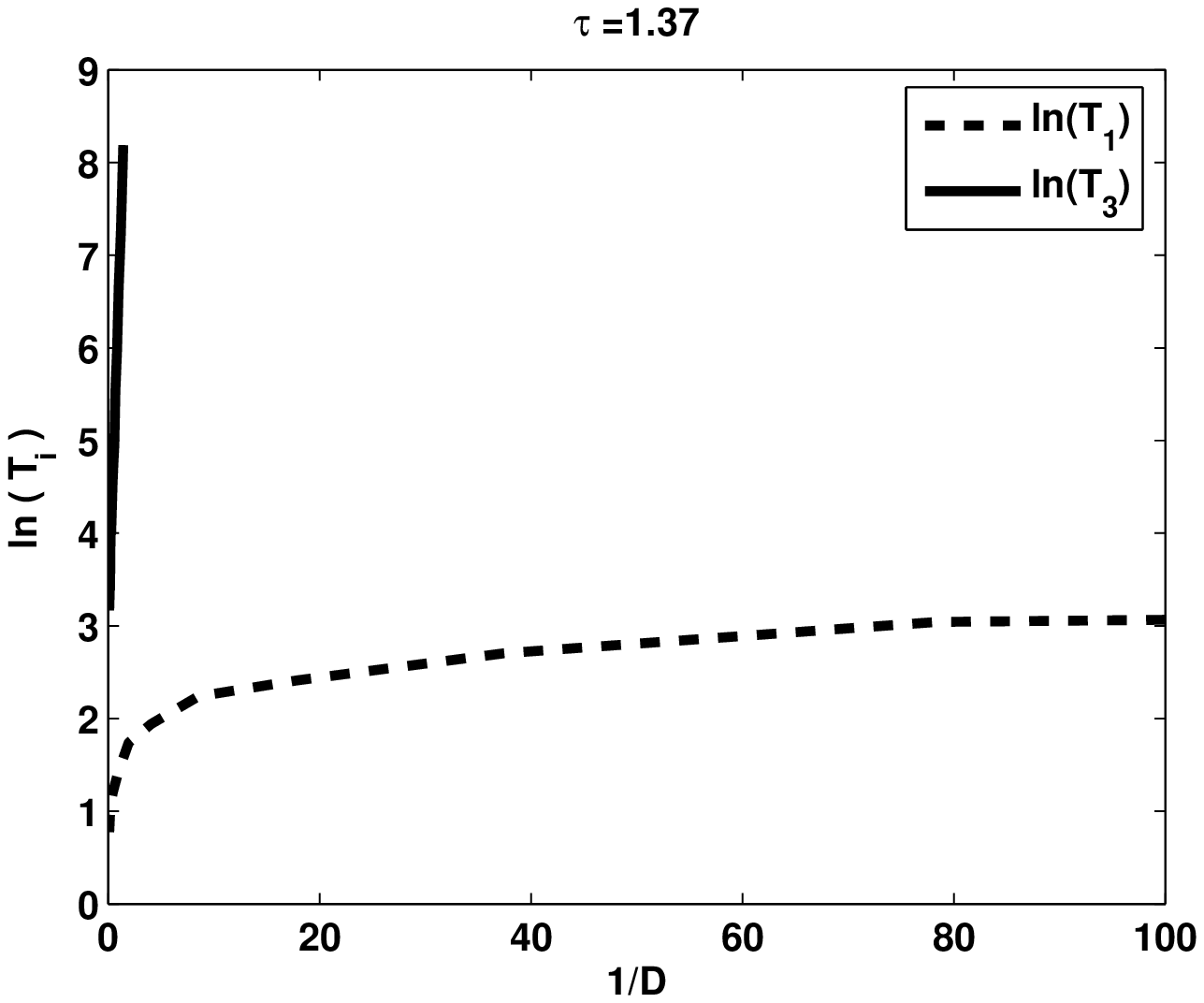}
\caption{\it
 Effect of the correlation time $\tau$ on the variation of the logarithmic average of the escape time from an attractor $A_3 \Longleftrightarrow A_1$ versus $1/D$, for the sets of parameters $S_7,$ shown in Table \ref{different} (where the frequencies of the two attractors are different).
The solid line ($T_3$) refers to escape from the outer circle, while the dashed line ($T_1$) refers to escapes from the inner circle.}
\label{fig:correlation2}
\end{center}
\end{figure}

We find that the pseudo-potential $U(A,D,\tau)$ depends on the amplitude of noise $D$ and on the correlation time $\tau$.
Also, the potential weakly (logarithmicly) depends upon the amplitude itself of the oscillations. However, in the low noise (and short correlation) limit  the term $Dln(A)/[2(1+\tau^2)]$ vanishes, thus indicating an Arrhenius-like escape time as in Eq.(\ref{Kramers}).
In the following Subsection we investigate the effects of correlation noise on the shape of the pseudo-potential $U$ of Eq.(\ref{eq20}) and we consider the cases where the frequencies $\Omega_1$ and $\Omega_3$ of the two stable attractors are identical and non-identical.

\begin{figure*}
\begin{center}
\includegraphics[height=5cm,width=7cm]{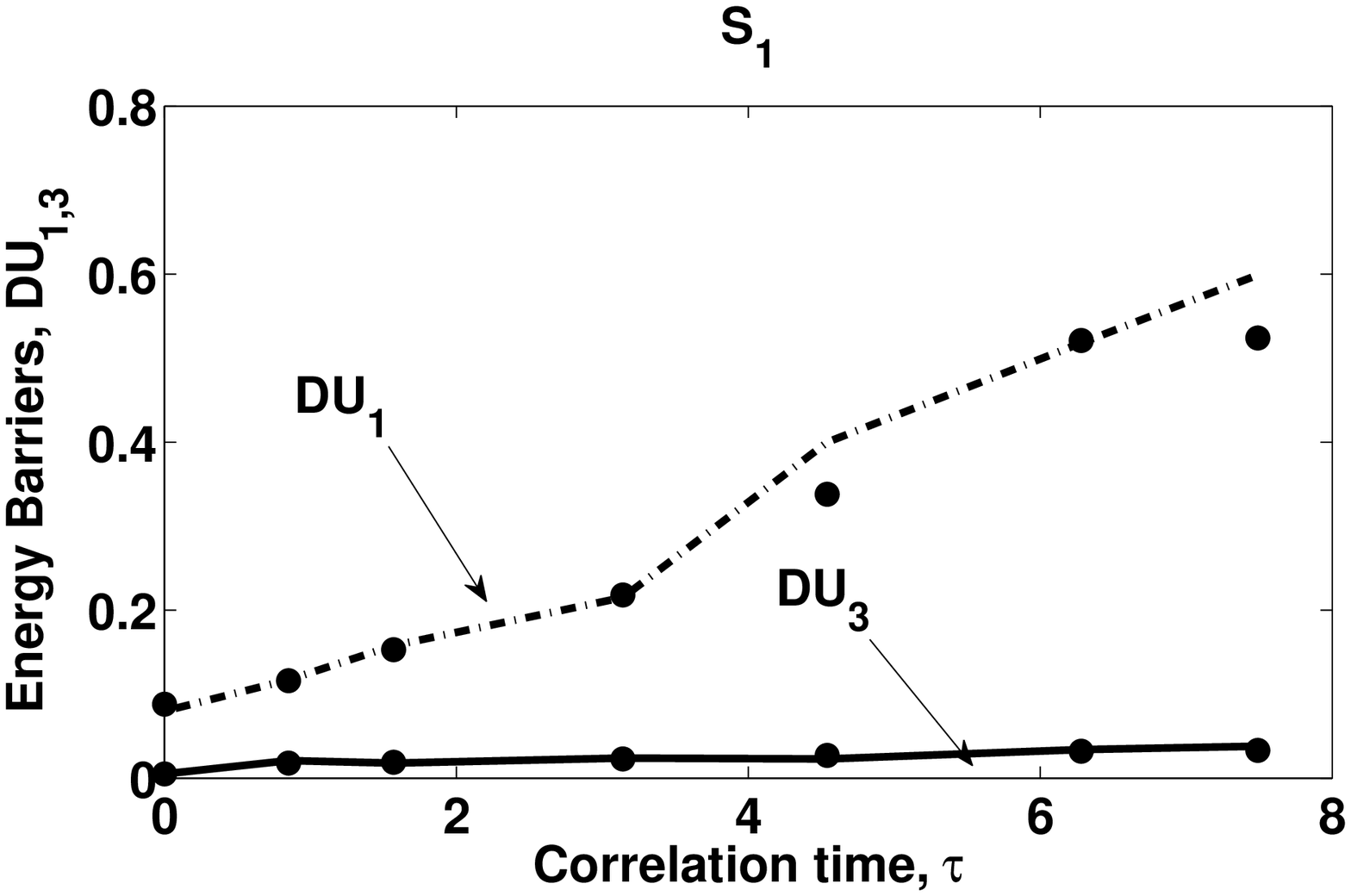}\hspace{1cm}
\includegraphics[height=5cm,width=7cm]{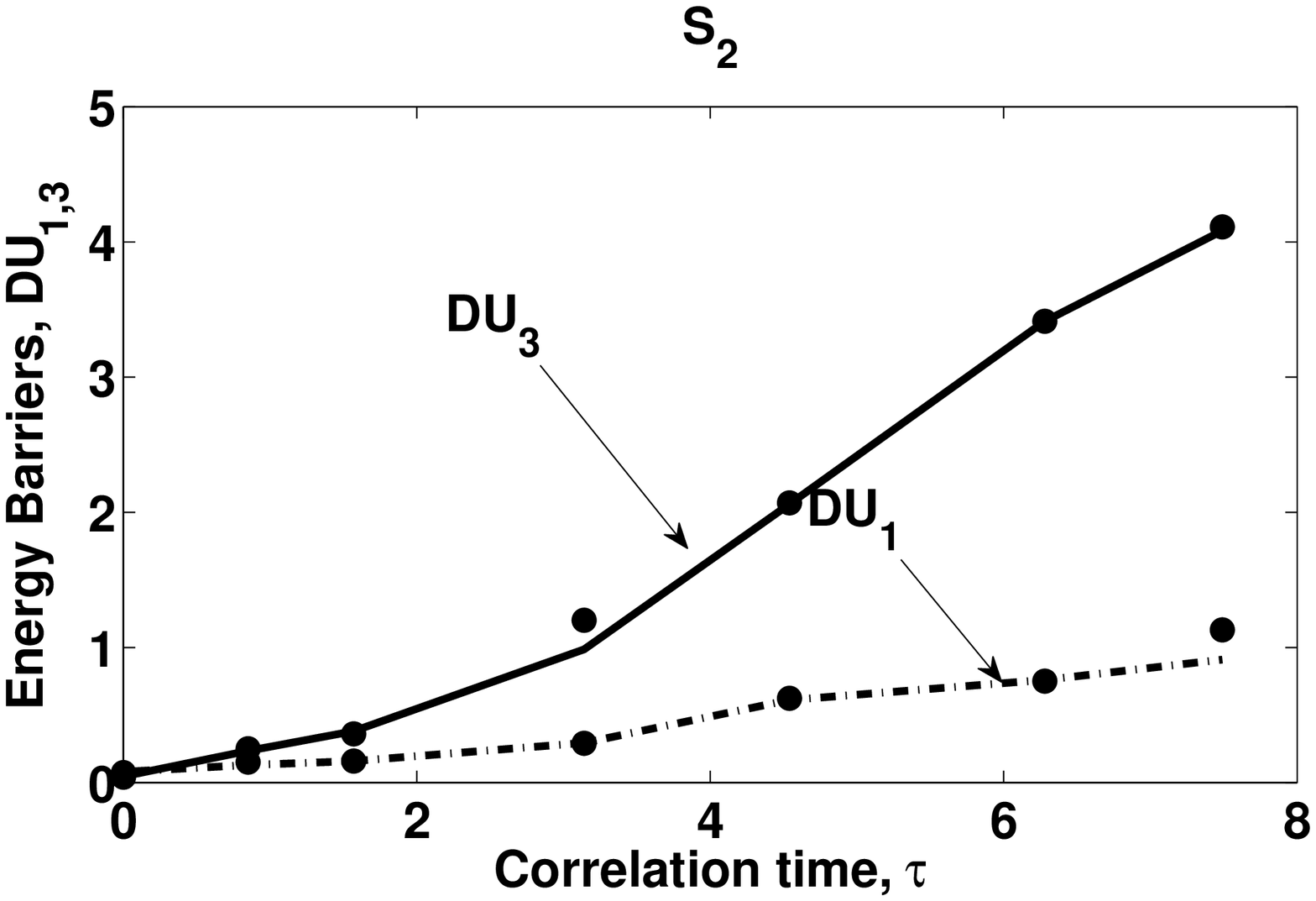}\\
\includegraphics[height=5cm,width=7cm]{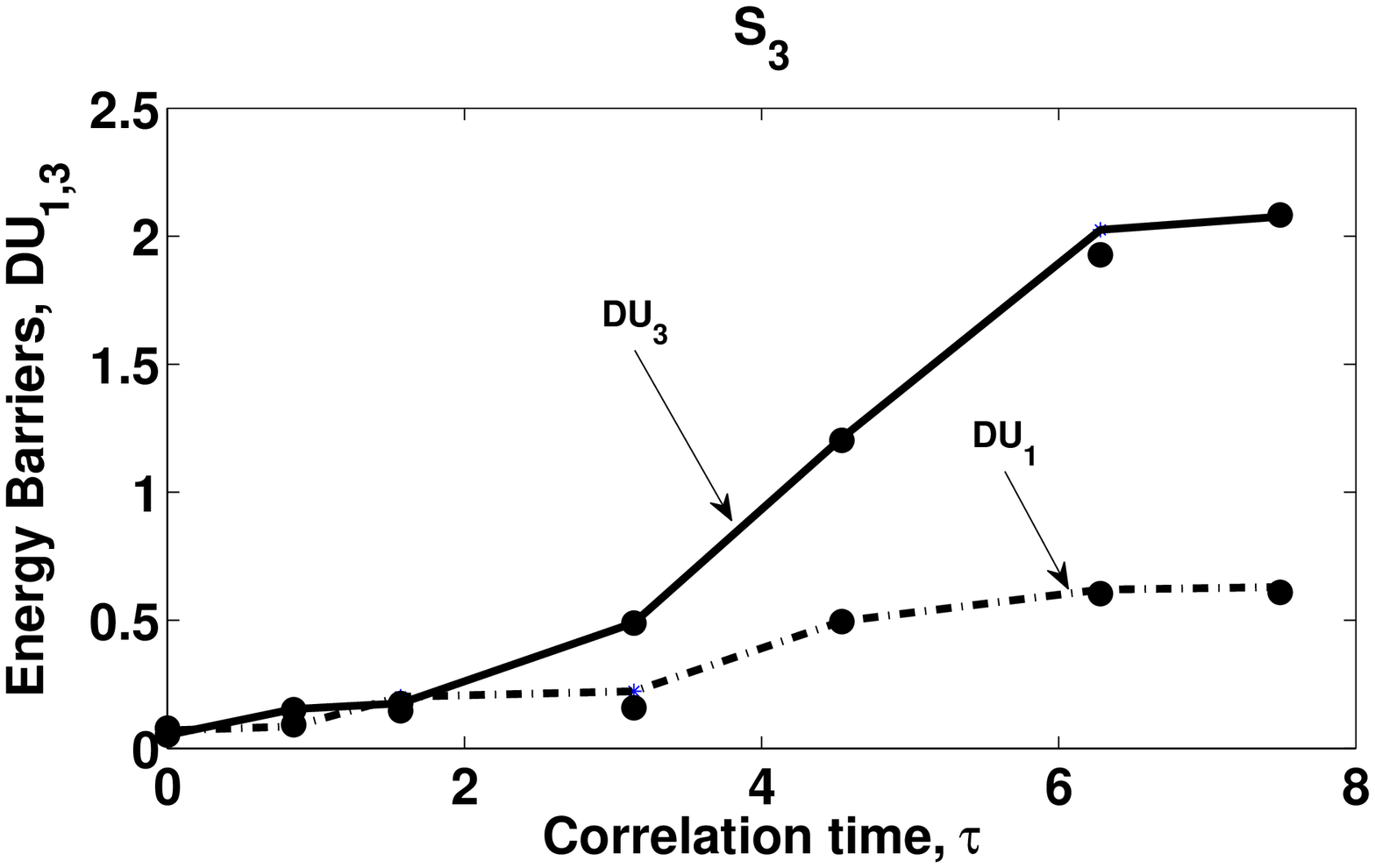}\hspace{1cm}
\includegraphics[height=5cm,width=7cm]{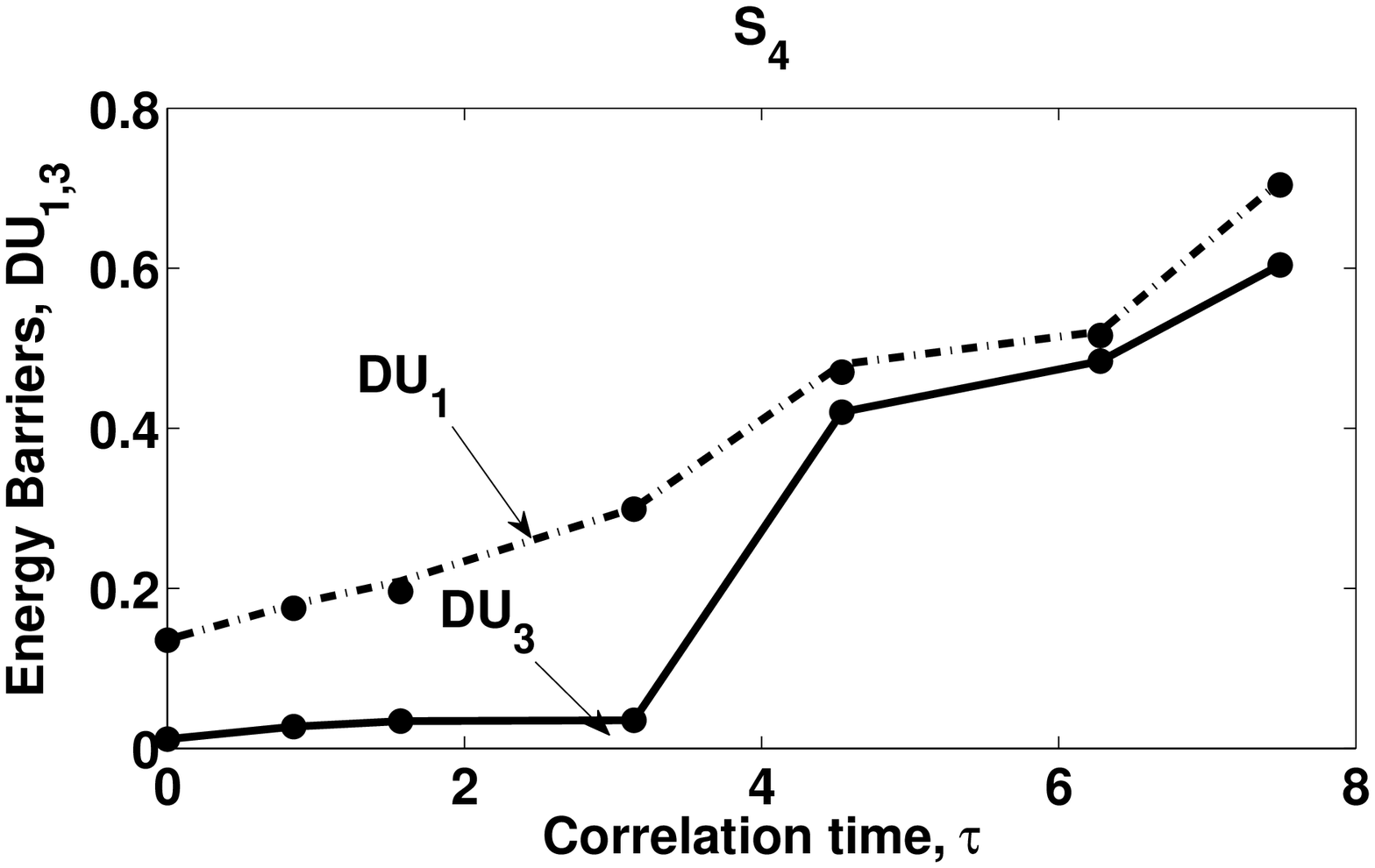}\\
\includegraphics[height=5cm,width=7cm]{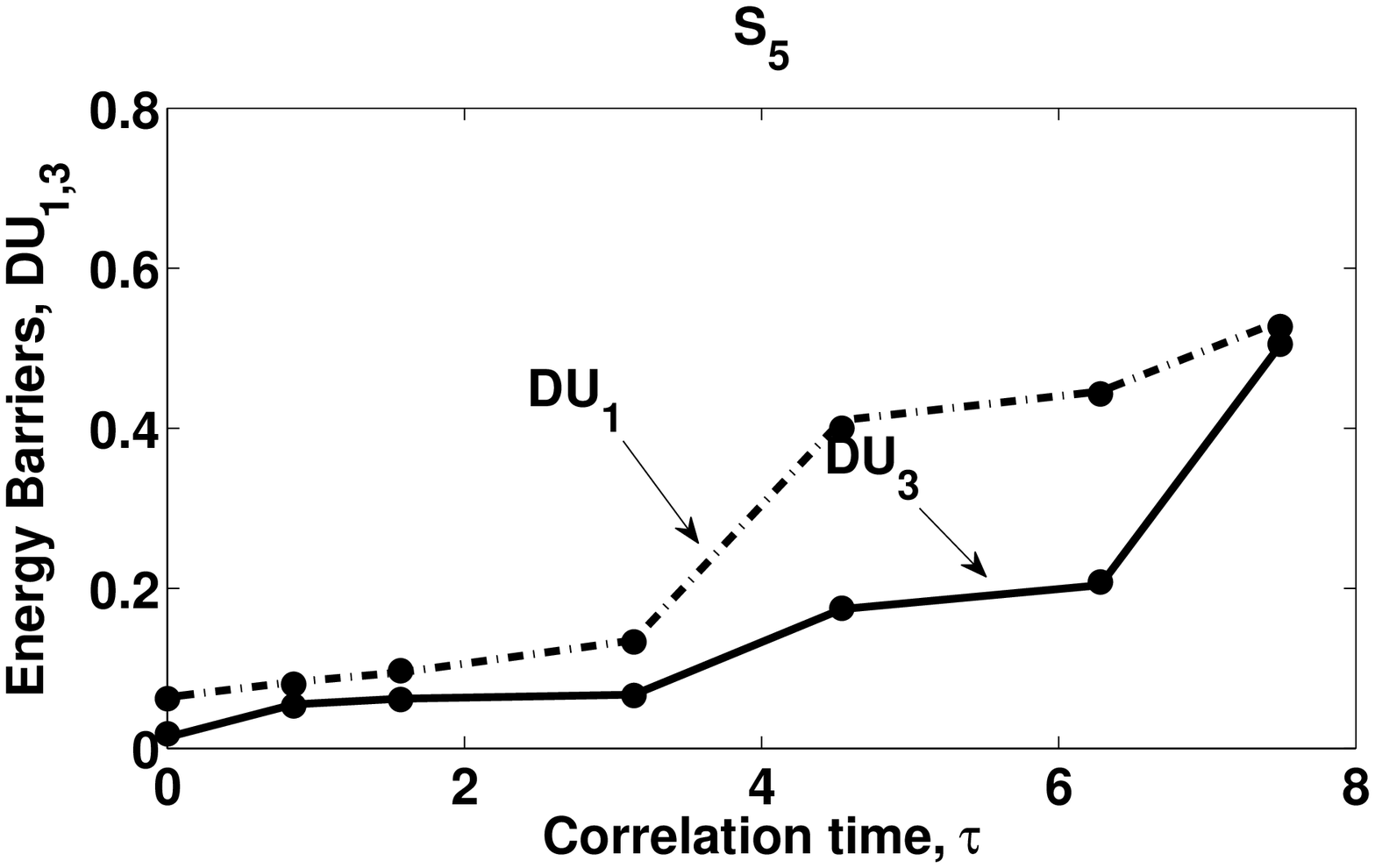}\hspace{1cm}
\includegraphics[height=5cm,width=7cm]{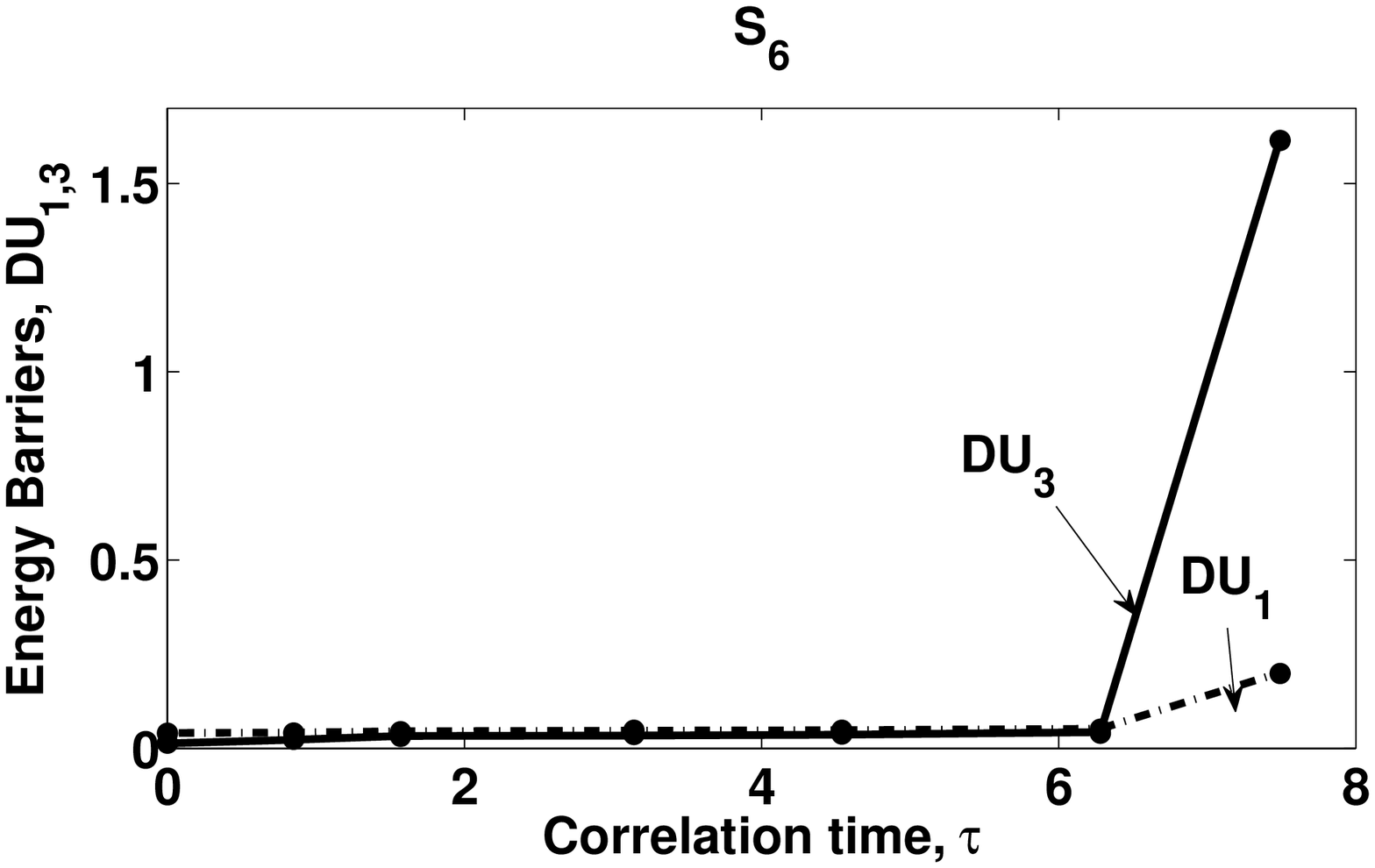}\\
\caption{\it
 Effective energy barriers versus the correlation time $\tau$ with the set of parameters $S_i$.
Circles (numerical) and solid line (estimate from stochastic average) correspond to escape from the outer cycle $A_3$, while stars (numerical) and dotted line (estimate from stochastic average) refers to escape from the inner cycle, $A_1$.
The parameters $\alpha$ and $\beta$ are the same as in Table \ref{identical}.
The frequencies of the two attractors are nearly identical.}
\label{fig:energy1}
\end{center}
\end{figure*}

\begin{figure}
\begin{center}
\includegraphics[height=5cm,width=7cm]{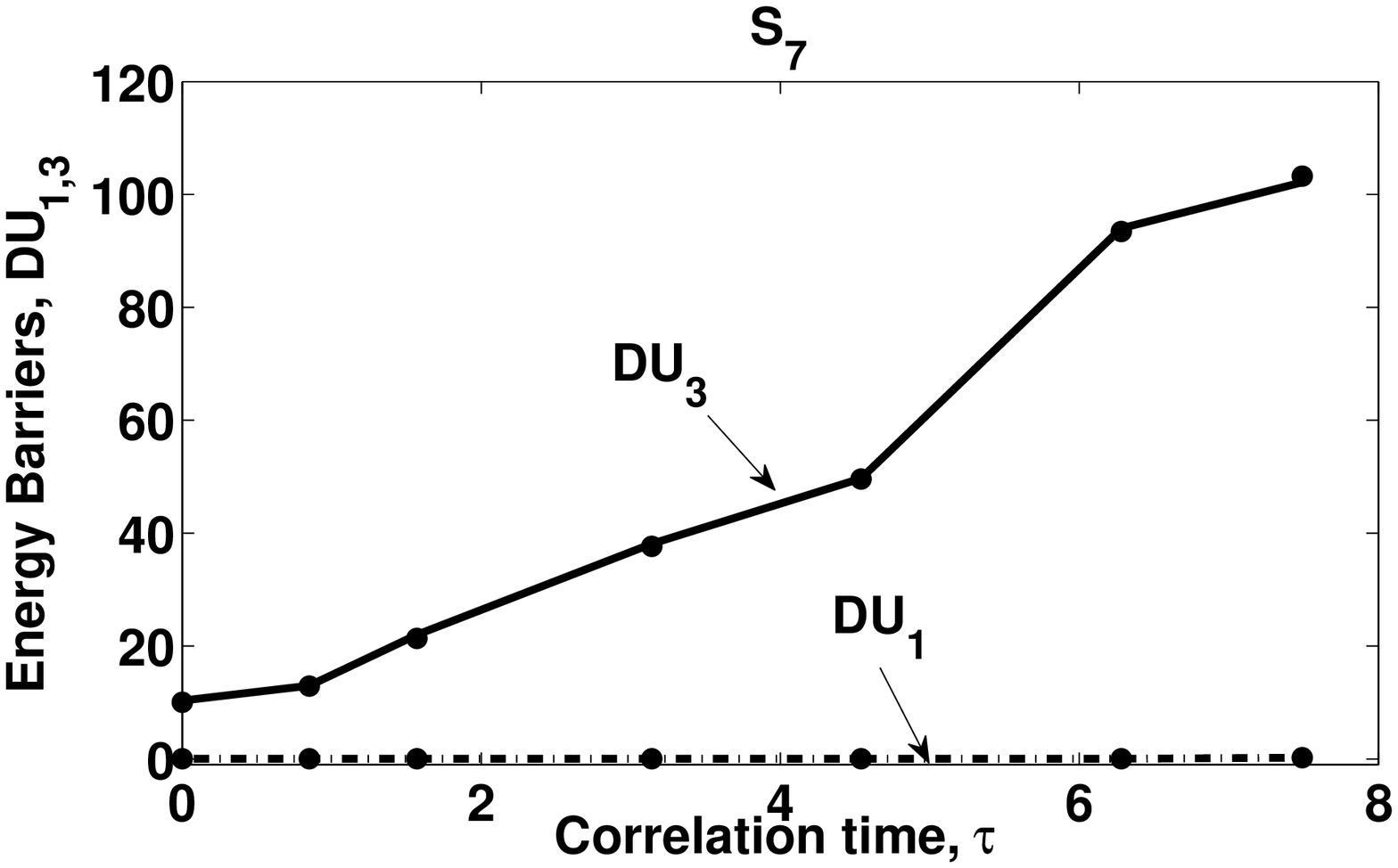}
\includegraphics[height=5cm,width=7cm]{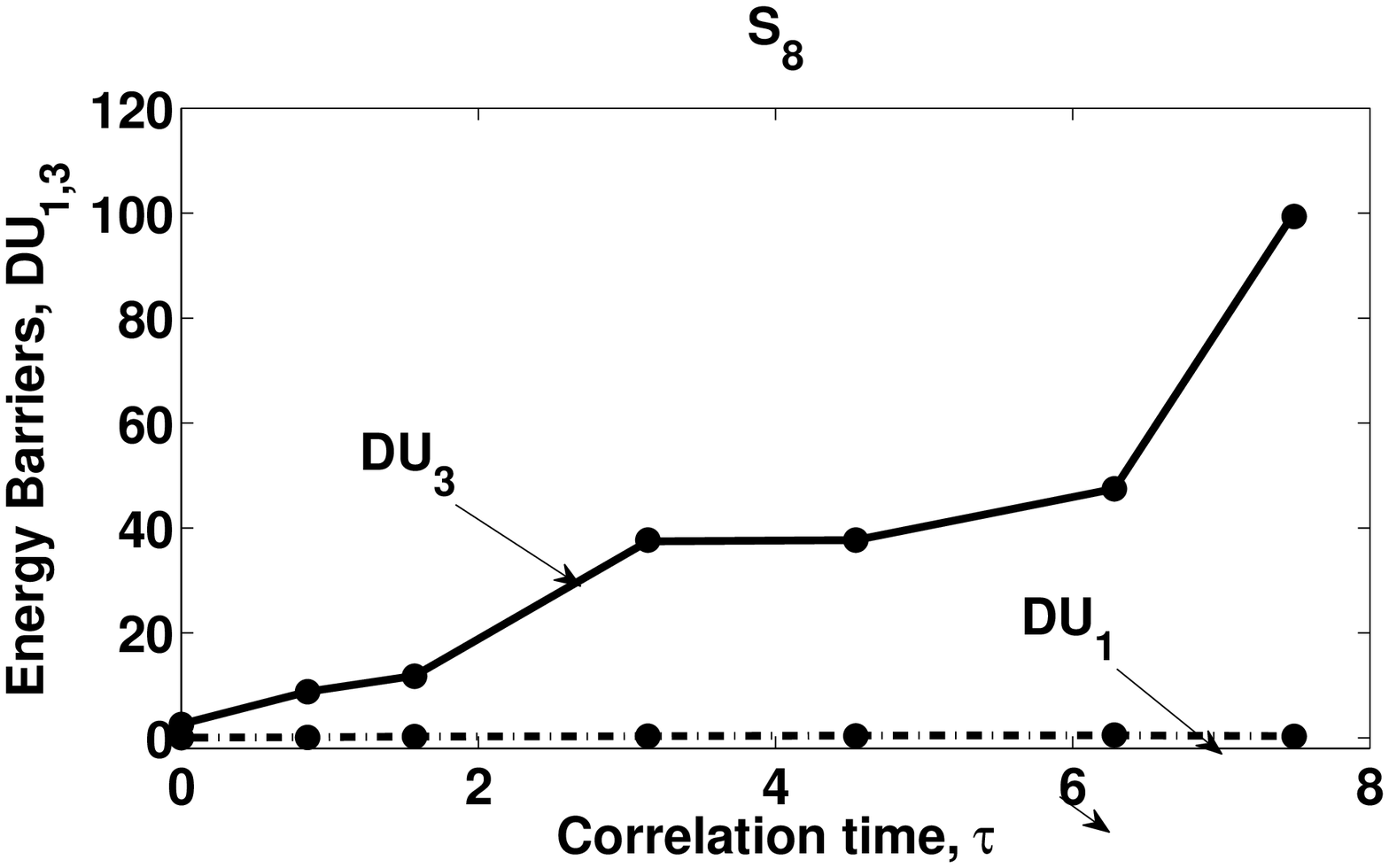}\\
\includegraphics[height=5cm,width=7cm]{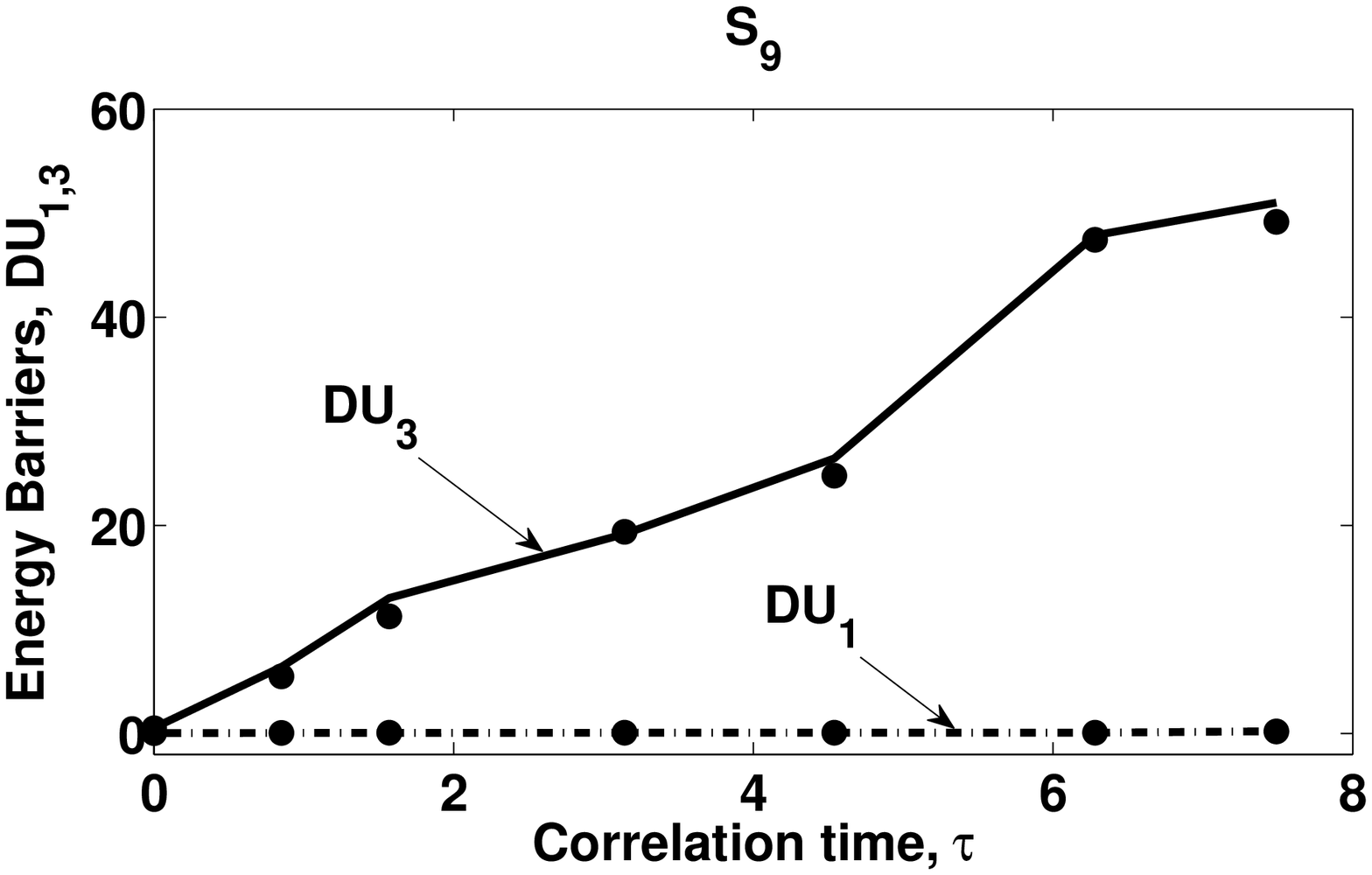}
\includegraphics[height=5cm,width=7cm]{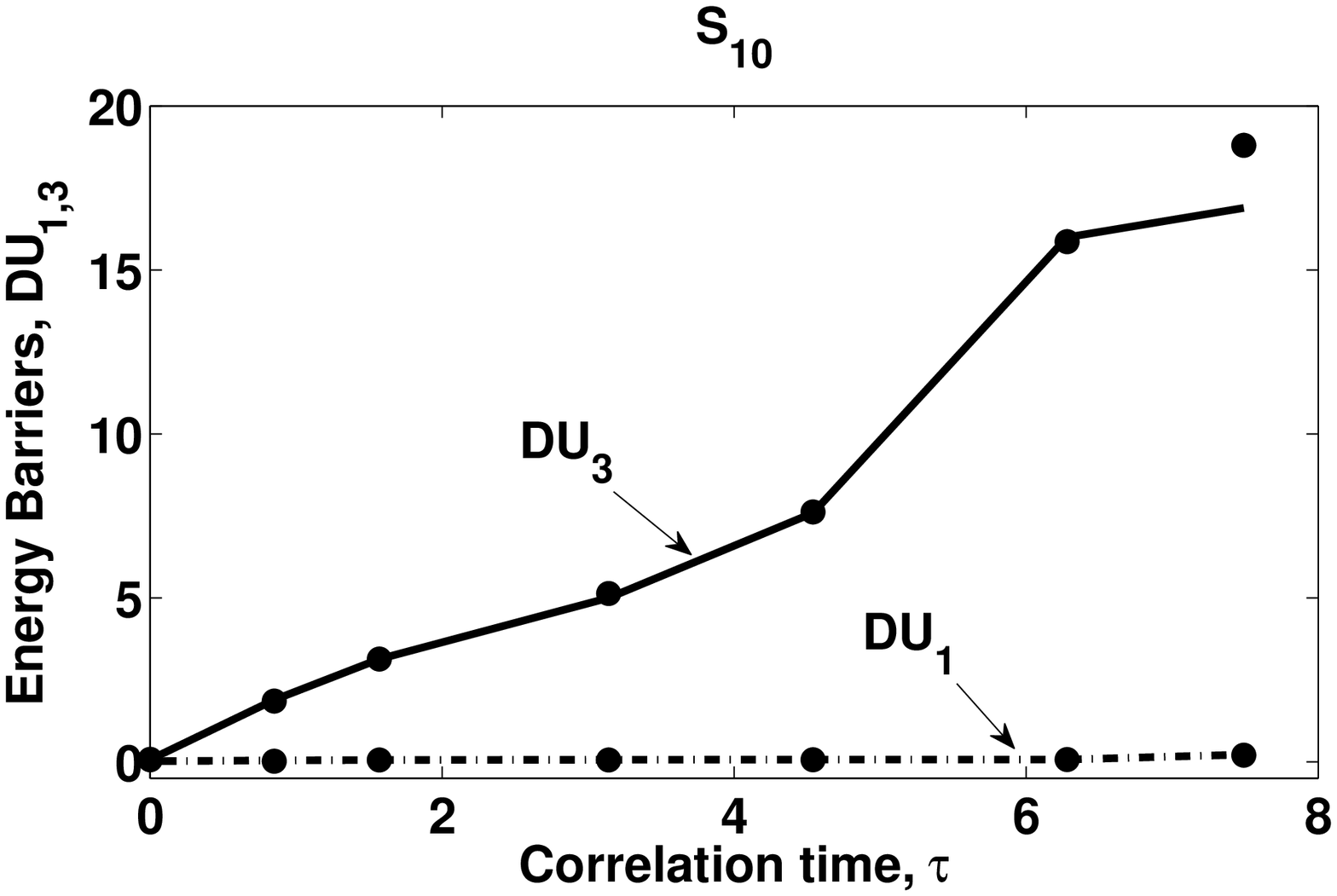}\\
\caption{\it
 Effective energy barriers versus the correlation time $\tau$ with the set of parameters $S_i$.
Circles (numerical) and solid line (estimate from stochastic average) correspond to escape from the outer cycle $A_3$, while stars (numerical) and dotted line (estimate from stochastic average) refers to escape from the inner cycle, $A_1$.
The parameters $\alpha$ and $\beta$ are the same as in Table \ref{different}.
The frequencies of the two attractors are different.}
\label{fig:energy2}
\end{center}
\end{figure}

\subsection{Influence of the correlated noise}
\label{influence}

In this Subsection we analyze the effects of correlated
noise on the shape of the pseudo-potential Eq.(\ref{eq20})
associated to the self-sustained birhythmic system (\ref{eq4}).
To do so, we analyze the effects of correlated noise on
the existence and the number of wells in the
pseudo-potential of the approximated averaged Eq.(\ref{eq19}).
We thus compare the noiseless shape of the pseudo-potential (\ref{eq20}) to the noisy case, keeping the correlation time constant.

We start with the case where the frequencies of the two attractors are almost identical, the sets of parameters $S_i(i=1,2,3,4,5,6)$ shown in the Table \ref{identical},  where the two frequencies are both equal to one i.e. $\Omega_1\simeq \Omega_3\simeq 1$.
This occurs, see Fig. \ref{fig:parameters}, for high values of the paramter $\beta$, close to the disappearance of the birhythmic region.

We display in Fig. \ref{fig:effects1} the variation of the pseudo-potential versus the amplitude $A$ for the sets of parameters $S_i$ and different values of the amplitude of the colored noise, $D$.  (For the free-noise self-sustained birhythmic system, \emph{i.e.} $D=0$, the shape of the pseudo-potential shows the double well  characterizing the birhythmic  behavior.)
We underline that the depth of the well embodies the stability of the system: the effective energy barrier becomes more efficacious when the well is deeper.
Increasing the noise amplitude, we find that the two wells persist, despite the effects of colored noise $D$, for attractors $S_1-S_6$.
However, stability is weakened, for the wells become more shallow.
The result is analogous to the effect of noise observed for uncorrelated noise \cite{Yamapi12}.

The response to correlated noise is similar also when the frequencies of the two attractors are different, {$\Omega_1 \neq \Omega_3$, or for the attractors $S_i (i=7,8,9,10)$ of Table \ref{different}, for the values of the  parameter $\beta$ are well below the disappearance of the birhythmic behavior, see Fig. \ref{fig:parameters}.

This is shown in Fig. \ref{fig:effects2}.
This behavior sets a limit for the highest noise intensity that preserves the birhythmic features.
The correlation, however, is low, $\tau = 0.01$; we conclude that noise with small correlation  time does not qualitatively change the overall picture.

For a more detailed analysis, in Figs. \ref{fig:correlation1} we show the average escape times as a function of the inverse noise intensity, slowly increasing the correlation time from very low values ($\tau = 0.001$) to $\tau = 2\pi$ (comparable with the period of the solution).
From the Figure it is clear that an Arrhenius-like behavior is conserved also for such values of the correlation. This is also confirmed in simulations of the birhythmic system with different natural frequencies, Figs. \ref{fig:correlation2}. We conclude that the effective energy barriers, or pseudo-potential, can be described by Eq. (\ref{Kramers}).
It also confirms that the qualitative features of the analysis are correct, for the solutions are of the Arrhenius type. From the slopes of the (log of the) escape times it is possible to numerically retrieve the pseudopotential behavior and to compare the analytic estimates of Eq.(\ref{eq20}) obtained through stochastic averaging.
This is done in Figs. \ref{fig:energy1} (for the case when the two attractors exhibit nearly the same frequency) and in Fig. \ref{fig:energy2} (when the two attractors are characterized by two distinct frequencies).
The effects of the correlation time on the effective energy barriers are provided in Figs. \ref{fig:energy1} and \ref{fig:energy2}.
In Fig. \ref{fig:energy1}, it is found that when the correlation time is very small, the pseudo-potential of the system remains qualitatively the same , as the energy barriers that are higher for zero correlation time stays higher for all explored values of the correlation time.
It is for both transitions are nearly identical, but increases with the variation of the correlation time. But in the case of the sets of parameters $S_{1,2,3,5}$,
the pseudo-potential becomes asymmetric, since the energy barriers of both transitions are different.
 It will be noted that even when the case $S_6$ is isolated  as the potential becomes asymmetric when the correlation time  approximates the period $T$ of the system.
 Figure \ref{fig:energy2} shows a similar transition as the pseudo-potential   which was originally symmetric, becomes asymmetric  with the variation of the correlation time. This  is also
 the case of the energy barrier $\Delta U_3$ for $A_3 \to A_1$ transition which increases
 with the variation of $\tau$, while $\Delta U_1$  does not change. 
 The two energy barriers  remain spaced from one on the other.
In both cases it appears that correlation tends to stabilize the pseudo-potential, as the two associated
wells become deeper. 
Also, the numerical findings (points) and the theoretical estimates (lines) are very close,
thus confirming the validity of  the stochastic averaging applied to Eq. (\ref{eq3}) also in the birhythmic region
in presence of correlated noise.

\section{Conclusions}
\label{conclusion}

The main objective of this work is to extend the concept of  pseudopotential (or quasipotential), that  has proved useful and reliable for uncorrelated noise, to a  birhythmic system affected by exponentially correlated noise.
We have shown that for a prototypal birhythmic system, a modified van der Pol oscillator, it is possible to derive with the stochastic  averaging method an effective Langevin equation that amounts to a particle trapped in a static potential.
The effective trapping energy Eq.(\ref{eq20}) is the main analytic result of this paper, that determines the stability  of the two attractors against large, noise-induced fluctuations, in the presence of exponentially correlated noise.
This results is consistent with the general finding that correlation increases the average escape time (e.g., \cite{Iwaniszewski96}).

However, the analysis relies on the assumption that the frequencies of the two solutions are equal (and coincide with the natural frequency of the linear oscillator, $1$), and that the frequency is independent of the amplitude of the oscillations.
These assumptions for the considered van der Pol system depends upon the parameters.
In particular the two frequencies are identical, within $1\%$, for some parameter values (see Fig. \ref{fig:parameters} and Table \ref{identical}), and grossly missed for some other parameter values (see of Table \ref{different}).
Therefore, the analytic results of the stochastic averaging have been compared with numerical simulations.
Interestingly, numerical simulations have confirmed the Arrhenius type behavior for the escape from the metastable orbits in the presence of correlated noise (that amounts to the existence of a quasi-potential for the dissipative system), both when the frequencies are similar and when the frequencies are different (Figs. \ref{fig:correlation1},\ref{fig:correlation2}).
We have thus confirmed that the three-dimensional system, Eqs.(\ref{eq3}), admits a pseudopotential \cite{Graham85} and that in the weak noise limit follows an Arrhenius-like behavior.
Moreover, numerical simulations have confirmed the quantitative accuracy of the stochastic averaging method, even when the two frequencies are non identical (see Figs. \ref{fig:energy1},\ref{fig:energy2}).

We conclude that an effective trapping energy can be employed to characterize the global stability properties of the birhythmic attractors, also in the presence of exponentially correlated noise.
An open question is why the stochastic average method is reliable when the two attractors are characterized by different frequencies.

\end{document}